\documentclass[usenatbib]{mnras}

\usepackage{multicol}
\usepackage{aas_macros,times,multirow,amsmath,amssymb,longtable,breqn}

\usepackage[T1]{fontenc}
\usepackage[varg]{txfonts}

\usepackage{grffile} 
\usepackage{graphicx}

\usepackage{color}
\usepackage{xcolor}

\def\ltsima{$\; \buildrel < \over \sim \;$}
\def\lsim{\lower.5ex\hbox{\ltsima}}
\def\gtsima{$\; \buildrel > \over \sim \;$}
\def\gsim{\lower.5ex\hbox{\gtsima}}

\newcommand{\be}{\begin{equation}}
\newcommand{\en}{\end{equation}}

\begin{document}
\label{firstpage}
\pagerange{\pageref{firstpage}--\pageref{lastpage}}

\title[Effects of magnetar bursts in pulsar wind nebulae]{Exploring the effects of magnetar bursts in pulsar wind nebulae}
\author[Martin, Torres, Zhang]
{J. Martin$^{1,2}$, D. F. Torres$^{1,2,3}$, Bing Zhang$^{4}$ \\
$^{1}$ Institute of Space Sciences (ICE, CSIC), Campus UAB, Carrer de Can Magrans s/n, 08193 Barcelona, Spain \\
$^{2}$ Institut d'Estudis Espacials de Catalunya (IEEC), Gran Capit\`a 2-4, 08034 Barcelona, Spain \\
$^{3}$ Instituci\'o Catalana de Recerca i Estudis Avan\c cats (ICREA), 08010 Barcelona, Spain \\
$^{4}$ Department of Physics and Astronomy, University of Nevada Las Vegas, NV 89154, USA
}

\date{}
\maketitle

\pubyear{2020}

\begin{abstract}
We explore possible effects of a magnetar burst on the radio, X-ray, and gamma-ray flux of a pulsar wind nebula (PWN). 
We assume that the 
burst injects electron-positron pairs or powers the magnetic field
and explore the total energy at injection and the spectral index needed in order to increase the X-ray flux by about one order of magnitude, as well as its decay time until reaching quiescence.
We also explore magnetically powered phenomenology that could explain a temporary increase of the PWN synchrotron emitted flux and perhaps the relatively common lack of PWNe surrounding magnetars.
This phenomenological study is of interest for fast radio bursts (FRBs) as well, given that the connection between magnetars and this kind of systems have been recently established observationally.

\end{abstract}


\section{Introduction}
\label{sec:intro}

Magnetar bursts are one of the most energetic phenomena in the Universe, with energies ranging between $10^{39}$ and $10^{47}$ erg at their emission peak and lasting between $\sim$0.1 and $\sim$40 s, see e.g., \citep{turolla2015}.
These outbursts are often accompanied by enhancements on the magnetar X-ray emission by factors between 10 and 1000 in timescales extending from a few weeks up to several years   \citep{cotizelati2018}.

Recent observations have revealed the existence of wind nebulae surrounding magnetars, as the one in Swift J1834.9-0846 \citep{Younes2016}.
Initial models suggested that this PWN was powered by a steady magnetic energy conversion \citep{Granot2017}.
But the appearance of adiabatic heating, being increasingly dominant as reverberation goes by, can be used to explain the nebula of Swift J1834.9-0846 only with a rotationally-powered injection \citep{Torres2017}.
In this scenario, magnetar nebulae are no different from normal PWNe, and,
follows on the discovery of low-field magnetars, e.g., \citep{Rea2010, Rea2014}, and of radio emission from magnetars, e.g.,  \citep{Camilo2006,Camilo2007,Anderson2012,Rea2012}.
However, how are such PWNe affected (if they are at all) when magnetars burst?

For instance, PSR J1119-6127 showed two short X-rays bursts, on 2016 July 27th \citep{kennea2016gcn} and 28th \citep{Younes2016gcn}, observed by the {\it Fermi}-Gamma-ray Burst Monitor and the {\it Swift}-Burst Alert Telescope, see also \citep{gogus2016}.
These bursts emitted a total energy of $3.7 \times 10^{38}$ and $5.2 \times 10^{38}$ erg just between 8 and 200 keV, and lasted 36 and 186 ms, respectively.
A PWN was already known to exist around J1119-6127 \citep{gonzalez2003}.
After analyzing new X-rays observations done with {\it Chandra}, \citep{safiharb2008} obtained an unabsorbed X-ray luminosity of $1.9 \times 10^{32}$ erg s$^{-1}$.
Three months after the outburst, \citep{blumer2017} measured again the PWN luminosity obtaining $1.9 \times 10^{33}$ erg s$^{-1}$ between 0.5 and 7 keV, an order of magnitude larger than the pre-burst luminosity.
Additionally, the photon index of the spectrum also changed from $\Gamma=1.2 \pm 0.8$ to $2.2 \pm 0.5$.
Something must have happened at the PWN.
However, the timing of the increase of the PWN flux is intriguing:
The maximum distance that the ejected particles could have traveled at the speed of light, assuming an 8.4 kpc distance to the pulsar, is about 2$^{\prime\prime}$ whereas the excess in flux has a radius of 10$^{\prime\prime}$ \citep{blumer2017}.
However, the distance to the pulsar can be overestimated, or the luminosity increase detected in the PWN may come from an earlier burst, or the relativistic particle injection of these magnetar events preceded the X-ray enhancement, or the ejecta could be beaming towards earth with a relativistic speed so as to promote a Doppler boosting.
There is no lack of {\it a priori} feasible options that would relate both events.

The idea that magnetars may produce relativistic particle outflows during an outburst was already hinted at by \citep{Thompson1996} and
explored in more detail by \citep{Harding1999}. 
The latter authors proposed that magnetars had episodic particle winds, with small duty cycle.
Also see \citep{Murase2016} for related studies with semi-analytical models, as well as the literature quoted below for related FRBs models.

It seems reasonable to think that in addition of photons, a pair plasma outflow is generated as a result of the bursts.
A short-lived nebula likely powered by the particles ejected during  magnetar bursts was found in the case of SGR 1900+14 \citep{Frail1999}.
Another radio source was also discovered following the giant flare of SGR 1806-20 \citep{Cameron2005,Gaensler2005}.
Copious relativistic particles must have been injected in order to generate these nebulae, these authors concluded.
Then, how does a magnetar burst affect a PWN if such is already surrounding the pulsar?
What observational signals can be expect from the PWN after the influence of the burst?
We here help to address these questions by considering a model for the magnetar J1834.9-0846 (although checked that our conclusions are generic and valid for other systems too). 
We analyze how its PWN could be affected by a magnetar burst, assuming either that the burst powers high-energy relativistic pairs injected into the PWN some time after, during, or before the burst; or an increase of the PWN magnetic field.
Before we proceed, we caveat upfront on the study limitations. What we present is phenomenological in nature: we shall not specify the mechanism of propagation of the perturbation nor which is the one for the possible acceleration of particles in any detail, although we offer deeper considerations below.
At the same time, injection cannot be really be instantaneously affecting all the PWN as we shall consider, but --depending on how the mechanism actually works-- may affect it only partially.
Despite these uncertainties, the intended exploration will hopefully fix the possible outcomes in very generic terms, and is thus deemed of use in exploring the possible effects of bursts in pre-existing PWNe.

\section{Burst injection on top of a steady PWN} \label{sec:par}

\subsection{Pre-burst spectrum}

\begin{table}
\scriptsize
\centering
\caption{Parameters used for the pre-burst spectrum of PWN J1834.9--0846. \label{tab:pwnpar}}
\begin{tabular}{lcc}
\hline
Parameter & Symbol & Value \\
\hline
{\it Pulsar parameters} \\
\hline
Age (yr) & $t_{age}$ & 8040\\
Braking index & $n$ & 2.2\\
Distance (kpc) & $d$ & 4\\
Initial spin-down age (yr) & $\tau_0$ & 193\\
Initial spin-down luminosity (erg s$^{-1}$) & $L_0$ & $4.67 \times 10^{38}$\\
\hline
{\it Injection parameters} \\
\hline
Energy break & $\gamma_b$ & $10^7$\\
Low energy index & $\alpha_l$ & 1\\
High energy index & $\alpha_h$ & 2.1\\
Containment factor & $\epsilon$ & 0.6\\
Magnetic fraction & $\eta$ & 0.045\\
\hline
{\it Background photon fields} \\
\hline
CMB temperature (K) & $T_{cmb}$ & 2.73\\
CMB energy density (eV cm$^{-3}$) & $w_{cmb}$ & 0.25\\
FIR temperature (K) & $T_{fir}$ & 25\\
FIR energy density (eV cm$^{-3}$) & $w_{fir}$ & 0.5\\
NIR temperature (K) & $T_{nir}$ & 3000\\
NIR energy density (eV cm$^{-3}$) & $w_{nir}$ & 1\\
\hline
{\it SNR parameters} \\
\hline
SN energy (erg) & $E_{sn}$ & $10^{51}$\\
Ejected mass (M$_\odot$) & $M_{ej}$ & 11.3\\
SNR density index  & $\omega$ & 9\\
ISM density (cm$^{-3}$) & $\rho_{ism}$ & 0.5\\
\hline
\end{tabular}
\end{table}

We use  {\sc TIDE} (see \citealt{Martin2012,Torres2014,Martin2016} for detailed discussions) to represent --as an example of a nebula surrounding a magnetar-- PWN J1834.9--0846. {\sc TIDE} couples the radiative properties of a time-dependent population of electrons, obtained by solving 
\begin{equation}
\label{diffloss}
\frac{\partial N(\gamma,t)}{\partial t}=Q(\gamma,t)-\frac{\partial}{\partial \gamma}\left[\dot{\gamma}(\gamma , t)N(\gamma,t) \right]-\frac{N(\gamma,t)}{\tau_{esc}(\gamma,t)},
\end{equation}
with a dynamical description of the PWN and the environment. The first term on the right hand side above represents the injection of particles, the second term accounts for the energy losses (synchrotron, inverse Compton --including self-Compton, Bremsstrahlung and adiabatic), and the third term accounts for escaping particles (we assume Bohm diffusion, defining the characteristic time scale $\tau_{esc}$).
For the steady injection in the PWN, we associate the particles to the spin-down power,
\begin{equation}
Q(\gamma,t)=Q_0(t)\left \{
\begin{array}{ll}
\left(\frac{\gamma}{\gamma_b} \right)^{-\alpha_1}  & {\rm for \;\;\;}\gamma \le \gamma_b,\\
 \left(\frac{\gamma}{\gamma_b} \right)^{-\alpha_2} & {\rm for \;\;\;}\gamma > \gamma_b,
\end{array}  \right .
\end{equation}
where the normalization term $Q_0(t)$ is computed via the spin-down luminosity of the pulsar $L_{sd}(t)$
\begin{equation}
(1-\eta)L_{sd}(t)=\int \gamma m_e c^2 Q(\gamma,t) \mathrm{d}\gamma,
\label{eqeta}
\end{equation}
and $\eta$ is  the instantaneous sharing parameter, describing the
distribution of the spin-down power into the nebula components.
The maximum energy that can be achieved is determined as the minimum 
between the gyroradius \citep{deJager2009} and the synchrotron limit \citep{deJager1996}.
The spin-down power evolves in time as 
\begin{equation}
\label{spindown}
L_{sd}(t)=L_0 \left(1+\frac{t}{\tau_0} \right)^{-\frac{n+1}{n-1}},
\end{equation}
where $n$ is the braking index, $\tau_0$ is the spin-down age (note that $\tau_0$ has a different definition than $\tau_{esc}$.), 
\begin{equation}
\tau_0 = \frac{2\tau }{ (n-1) } - t_{age}
\end{equation}
and $L_0$ is the initial spin-down power
\begin{equation}
L_0=L_{sd} \left(1+ \frac{ t_{age}}{\tau_0}\right)^{\frac{n+1}{n-1}}.
\end{equation}

Note that here we are only considering that all particles injected into the PWN are accelerated through a transfer of energy from the spin-down of the magnetar, we do not use the magnetic field decay as an energy reservoir. Then, it is difficult from our model to make constraints associated with the magnetic field decay as of now, although this may constitute an interesting direction for further future work.
However, we can already make an estimate of the influence of the magnetic field evolution if we were to incorporate the magnetic field decay in our model by using, for instance, the bottom plot of figure 3 in the work of \citet{Beniamini2019}. 
From Eq. \ref{spindown} above, we see that most of the energy is released during the first $\tau_0$ years (spin-down initial age). In particular, for the dipolar model ($n=3$),
\begin{equation}
E=\int_0^{\tau_0} L_0 \left(1+\frac{t}{\tau_0} \right)^{-2} \mathrm{d}t=\frac{L_0 \tau_0}{2}
\end{equation}
one half of the total energy is released then.
From figure 3 of \citet{Beniamini2019}, we see that most of the rotational energy is released in a period of time that depends on the ratio between the magnetic field decay timescale $\tau_B$ and the spin-decay timescale $\tau_\Omega$ (equivalent to $\tau_0$ in our model). 
When $\tau_B < \tau_\Omega$, the magnetic field decay timescale determines the moment when the rotational energy losses decays significantly, keeping the period constant. 
On the other hand, when $\tau_B > \tau_\Omega$, it is the spin decay timescale the one determining such moment. 
If we consider that $\tau_B$ is approximately 10 kyr \citep{Goldreich1992,Beniamini2019}, this quantity is roughly two orders of magnitude higher than the initial spin-down age that we are considering here (and usually larger by typically a factor of 3 when compared to initial spin-down age of other pulsars, see e.g. \citealt{Torres2014}).
Thus, we deduce that the magnetic field decay timescale is not expected to influence significantly the evolution of the spin-down luminosity in our case, and we shall neglect it for the moment.
Further analysis of the evolution would be needed to gather more insights.

To complete the model we consider a nebular magnetic field
that is also powered by the spin-down, and is subject to adiabatic losses,
\begin{equation}
\frac{d W_B(t)}{dt}=\eta L_{sd}(t)-\frac{W_B(t)}{R(t)} \frac{d R(t)}{dt},
\label{eq:magfield}
\end{equation}
where $W_B=(B^2/8 \pi) (R^3 \, 4\pi/3)$ is the total magnetic energy,

The parameters for J1834.9--0846 shown in Table \ref{tab:pwnpar} are essentially the same as the ones used in \cite{Torres2017}, and so 
is the resulting spectral model for the PWN that was found to agree with the X-ray measurements.
Here, there is only a small shift ($<100$ years) in the age of the PWN related with a refinement in the PWN dynamics across the reverberation process (see \cite{Martin2020} for details), which in the context of other uncertainties is irrelevant. The dynamics results from solving
\begin{eqnarray}
\frac{dM(t)}{dt} & = & 4 \upi R^2(t) \rho_{ej}(R,t)[v(t)-v_{ej}(R,t)]\\
M(t)\frac{dv(t)}{dt} & = & 4 \upi R^2(t) \left[P_{pwn}(t)-P_{ej}(R,t) \right. \nonumber \\
 & & \qquad \left.-\rho_{ej}(R,t)\left(v(t)-v_{ej}(R,t) \right)^2 \right] \label{eqmom}
\end{eqnarray}
during the free expansion phase, and 
\begin{eqnarray}
\frac{dM(t)}{dt} & = & 0 \nonumber \\
M(t)\frac{dv(t)}{dt} & = &4 \upi R^2(t) [P_{pwn}(t)-P_{ej}(R,t)]
\end{eqnarray}
when PWN shell interacts with the reverses shock of the SNR.

Next we shall consider (without specifying which one is it) that a magnetar burst promotes a mechanism by which particles, or field,
or both, are deposited in the surrounding PWN, producing emission on top of the pre-burst spectrum. 
For instance, in a picture similar to GRBs, the magnetar flare could be associated with an outflow that carries a kinetic and magnetic energy. 
This magnetized shell may collide with the PWN, driving a forward shock and accelerating electrons to higher energies.
The magnetic field in the forward shock region would also be enhanced with respect to the original PWN field, e.g., due to shock compression and/or Weibel instability in the shock downstream. 
Depending on the magnetization of the shell, there may be also a reverse shock entering the shell. 
If it exists, the reverse shock could also accelerate particles in a stronger magnetic field. 
Both the FS and RS can power brighter emission than the original PWN emission.

\subsection{Burst energetics into particle injection}

\begin{figure*}
\centering
\includegraphics[width=0.45\textwidth]{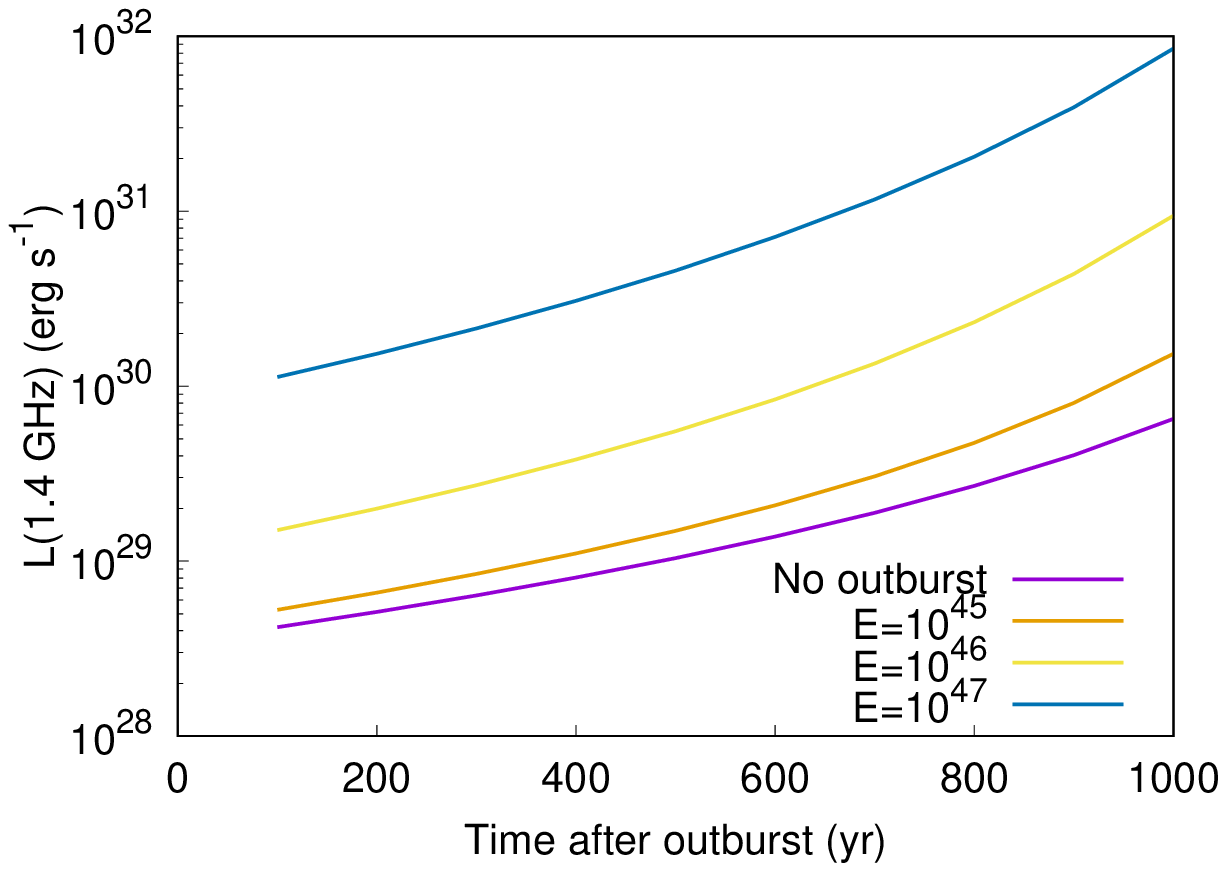}
\includegraphics[width=0.45\textwidth]{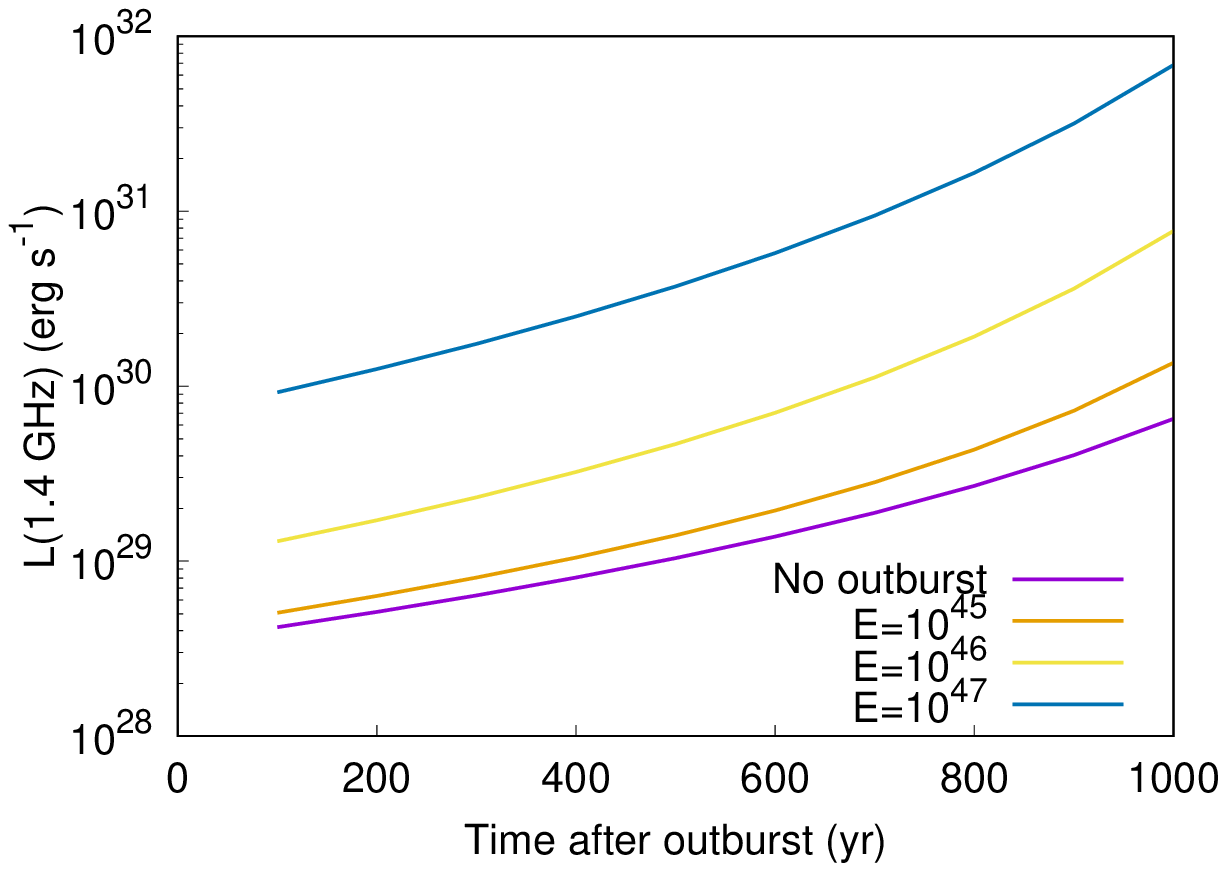}
\includegraphics[width=0.45\textwidth]{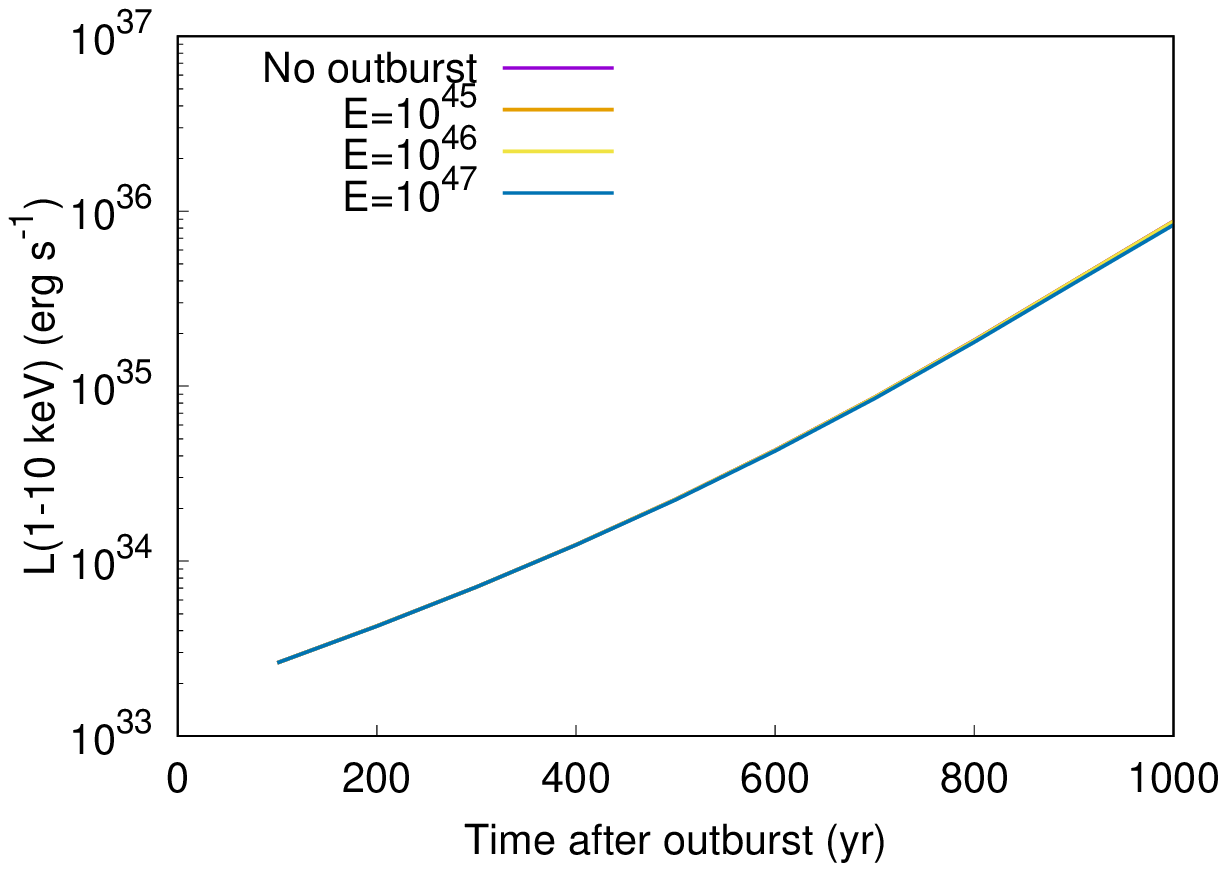}
\includegraphics[width=0.45\textwidth]{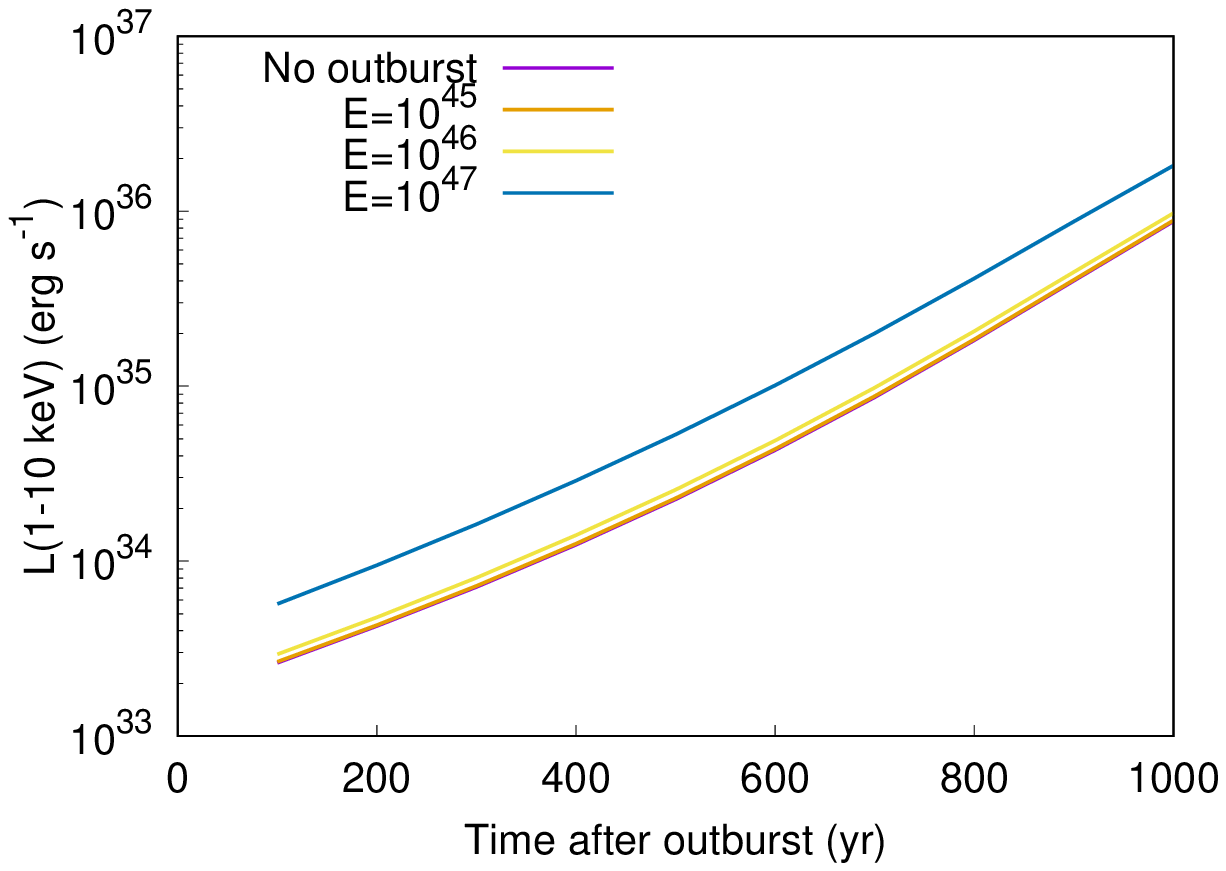}
\includegraphics[width=0.45\textwidth]{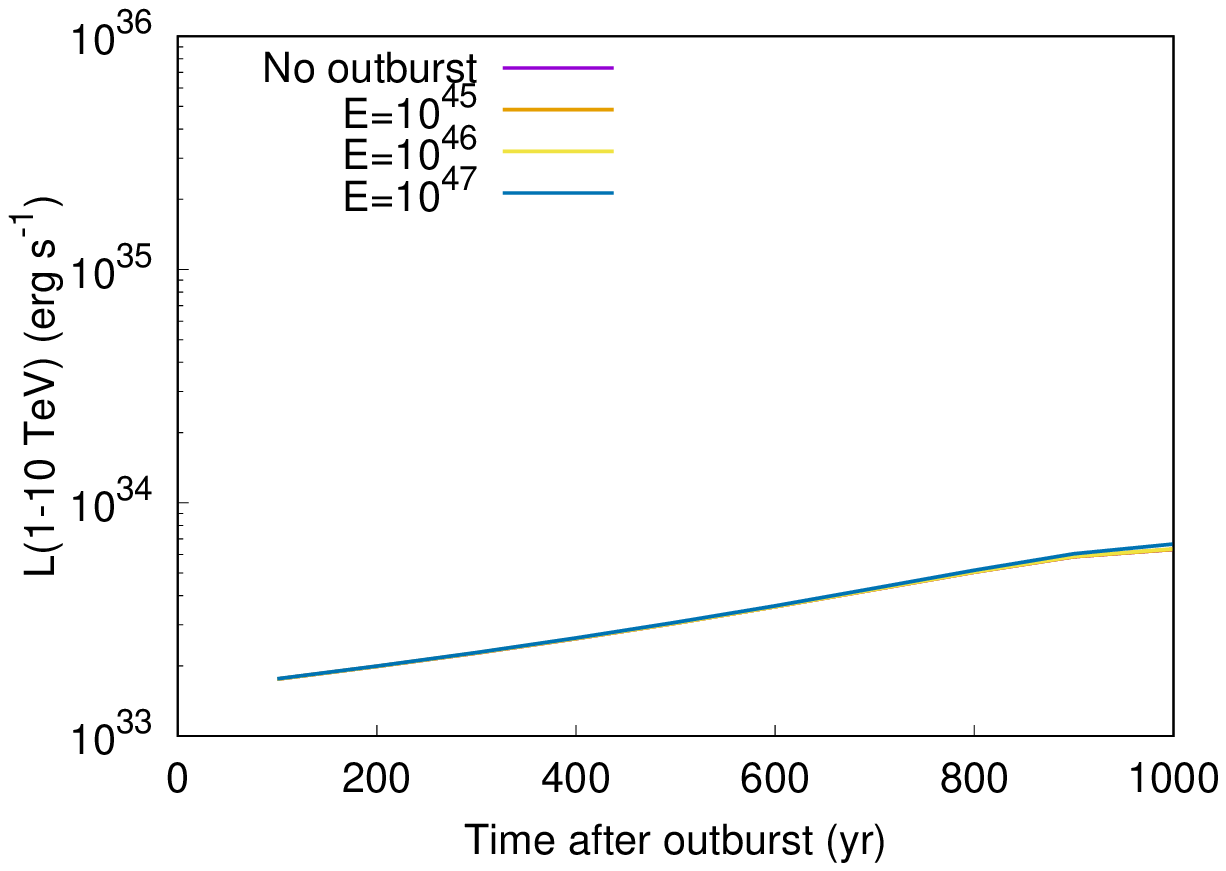}
\includegraphics[width=0.45\textwidth]{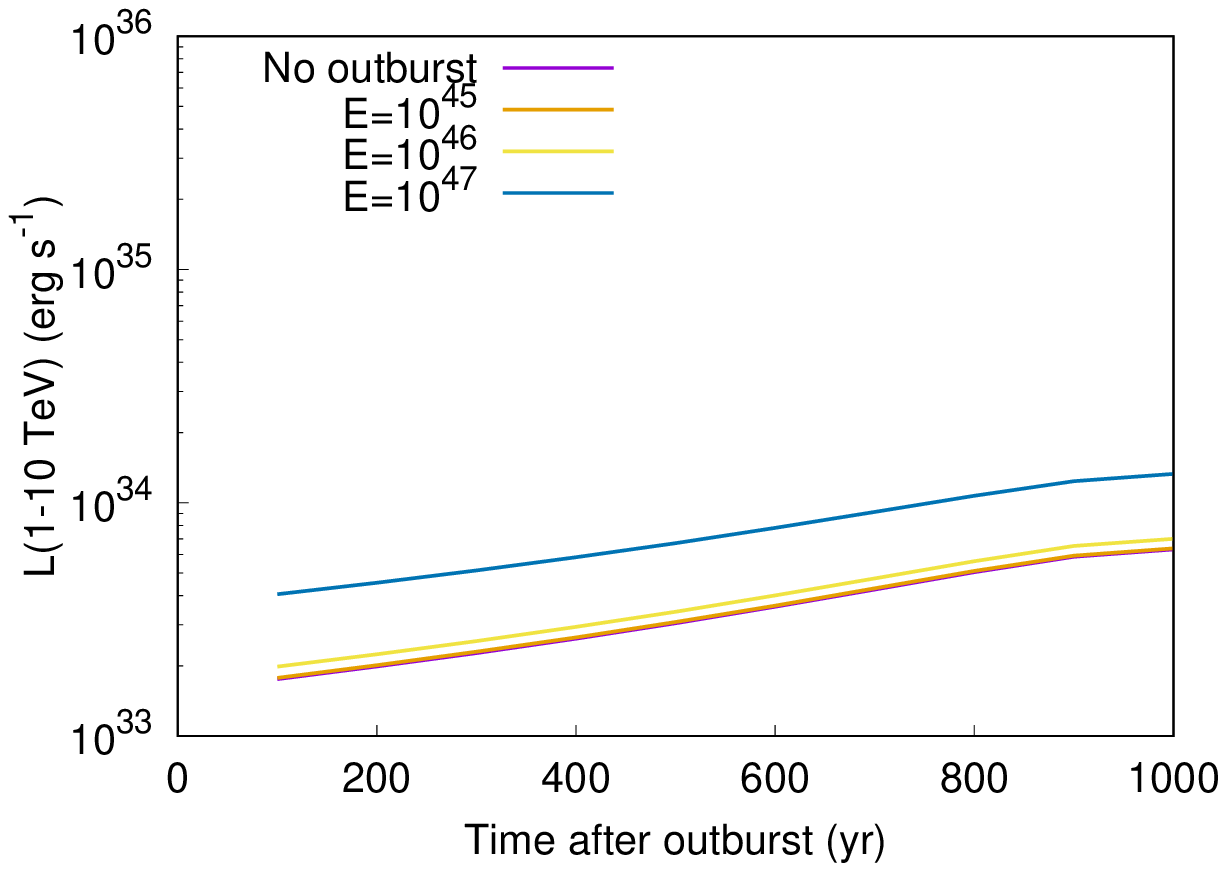}
\caption{Evolution of the luminosity in radio (1.4 GHz, top panels), X-rays (1-10 keV, middle panels), and VHE gamma-rays (1-10 TeV, bottom panels) in the 'burst energetics into particle injection' case.}
\label{fig:lumevolradio}
\end{figure*}

Let us first consider that a significant amount of relativistic particles is injected by a putative magnetar burst.
This happens roughly instantaneously in comparison with the dynamical timescales of the PWN.
Thus, if travelling at the speed of light, these particles will be reaching the termination shock of the PWN some months later, roughly at the same time.
The termination shock position is approximately given by (e.g., \citealt{Gelfand2009}),
\begin{equation}
R_{ts}=\sqrt{\frac{L(t)}{4 \upi \chi c P_{pwn}(t)}},
\end{equation}
where $L(t)$ is the spin-down luminosity of the pulsar, $P_{pwn}$ is the internal pressure of the PWN and $\chi$ is the filling factor, which is equal to 1 for an isotropic wind.
In the case of J1834.9--0846, the termination shock is then located at $\sim 0.05$ pc.
Particles injected from the center of the PWN travelling at the speed of light would reach the termination shock, at the base of the PWN, in $\sim$2 months.
Thus, in order to fix a concrete example with matching energetics, 
we shall consider an injection of particles during 1 second, with a total energy $E_{out}$ of $10^{45}$, $10^{46}$ and $10^{47}$ erg.
If a signficant portion of this energy goes into detectable radiation, we would be seen a giant flare. We choose such range in order to see the effects in the lightcurves shown in Figures \ref{fig:lumevolradio}, \ref{fig:lummagevol} and \ref{fig:elecmag} better.
This injection overcomes the steady injection from the pulsar spin-down, as described above, 
which may in fact be completely absent during the time of the burst.

For the particle distribution resulting from the burst, we shall assume 
an {\it exponentially-cut power-law }
\begin{equation}
{\cal I}(\gamma)=A \gamma^{-\delta} e^{-\gamma/\gamma_c}, \hspace{.15cm} {\rm such \,\, that} \hspace{.15cm}
E_{out}=\int
\gamma m_e c^2 {\cal I} (\gamma) \mathrm{d}\gamma
\end{equation}
being $\delta$ the spectral distribution index, $\gamma_c$ the cutoff energy, and $A$ the normalization constant. 
This distribution could result from particles that were  injected by the burst, but contained by the termination shock, re-isotropized, and re-accelerated there; or be the particle distribution directly injected by the burst (accelerated closer to the pulsar), that overcoming the termination shock.
We shall adopt $\delta=$1.5, 2.0 and 2.5.
Given that we have observational constraints on the burst, even for the same source and event \citep{Younes2016gcn}, as well as for the emission at both GeV \citep{Li2017} and TeV \citep{Aleksic2013b}, that imply that magnetars are not bright gamma-ray sources (albeit their PWN might be), $\gamma_c$ cannot be too high.
We adopt $\gamma_c=1$ TeV and 30 TeV.

Table \ref{tab:lumpar} shows the X-ray and VHE luminosities 1 year after the burst, considering that all the energy is invested in accelerating particles.
Only the spectrum in radio increases in all the cases explored.
X-ray and VHE emission is only affected in the more extreme cases ($E_{out}=10^{46}$ and $10^{47}$ erg) with a hard particle spectrum ($\delta \le 2$) and particles accelerated up to 30 TeV.
For $E_{max}=1$ TeV, particles have not enough energy to contribute significantly  (the low magnetic field of the PWN, 4.4 $\mu$G, inferred from the model, plays a role in this conclusion, which could be different in other PWNe).

When particles reach the termination shock, the luminosity is increased almost instantaneously (by a factor $\sim$20-30 in radio and $\sim$10-20 in X-rays and VHE in the most extreme cases).
The loss timescales for particles are too large in order to see a decay in the simulation period.
Actually, once the luminosity is enhanced, it remains constant along the 1-year simulation.
This fact is easy to understand considering the synchrotron timescale for PWN J1834.9-0846.
For particles with enough energy to radiate in X-rays ($\gamma \sim 10^8$), the timescale to lose all their energy is given by
\begin{equation}
\tau_{syn} \simeq 245.37 \left(\frac{1\ \mu\text{G}}{B} \right)^2 \left(\frac{10^8}{\gamma} \right)\ \text{kyr}.
\end{equation}
For a particle with $\gamma=10^8$ in a magnetic field of 4.4 $\mu$G,  the time required to lose all its energy through synchrotron radiation is $\sim$13~kyr.

Fig. \ref{fig:lumevolradio} shows the long-term evolution of the luminosities for the case $\delta = 2.0$ and different energetics. Results are similar for other values of $\delta$.
Although we let evolve the system for 1 kyr, the difference between the luminosity after the outburst in comparison with the no-outburst case diminishes very slowly.
And in the radio range, it actually increases due to the the compression of the PWN.

\subsection{Burst powering of the magnetic field, normal decay}
\label{secmag}

We now explore the possibility of having an increase of the mean magnetic field of the PWN due to a direct injection of energy as a result of the  burst; i.e., we consider
\begin{equation}
\frac{d W_B(t)}{dt}=\eta L_{sd}(t)+\frac{E_{out}}{t_{out}}-\frac{W_B(t)}{R(t)} \frac{d R(t)}{dt},
\end{equation}
being $t_{out}$ the duration of the burst.

\begin{figure}
\centering
\includegraphics[width=0.45\textwidth]{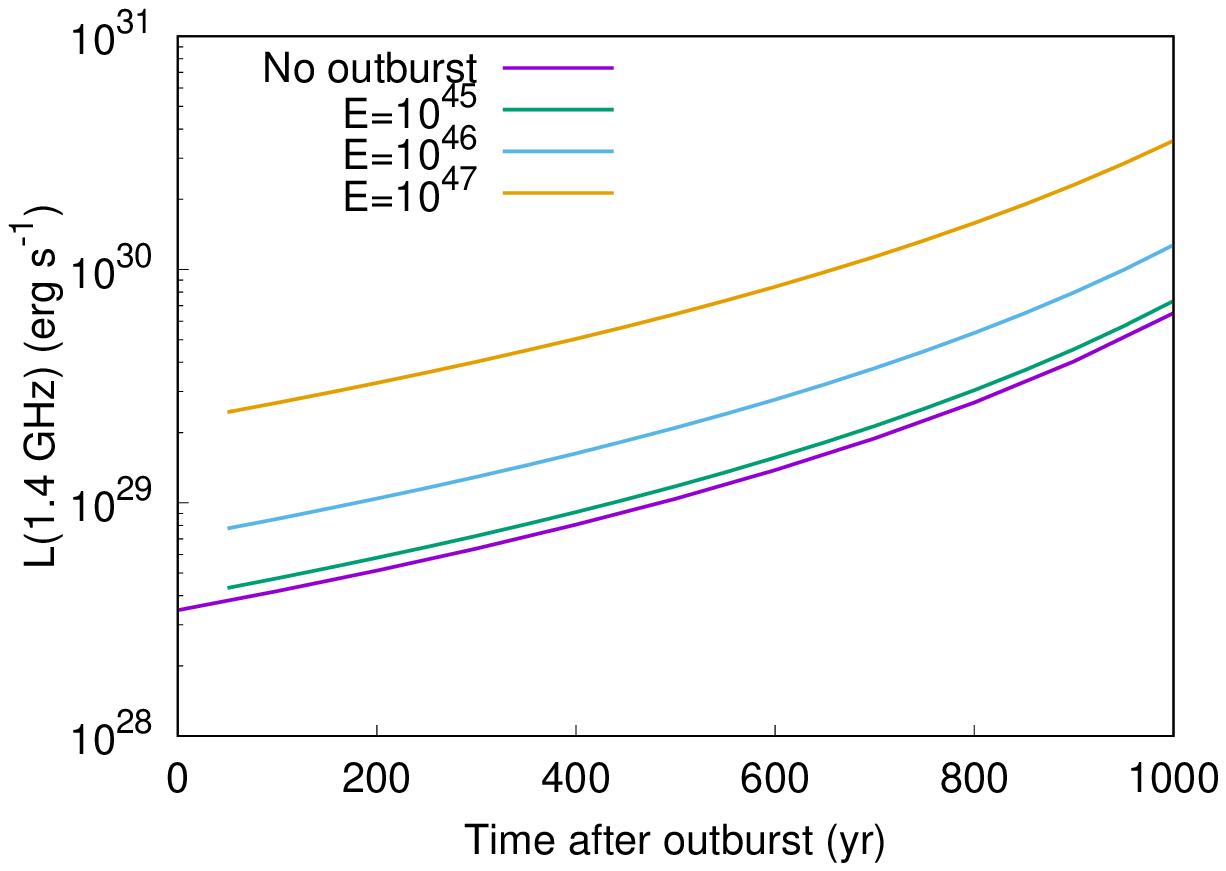}
\includegraphics[width=0.45\textwidth]{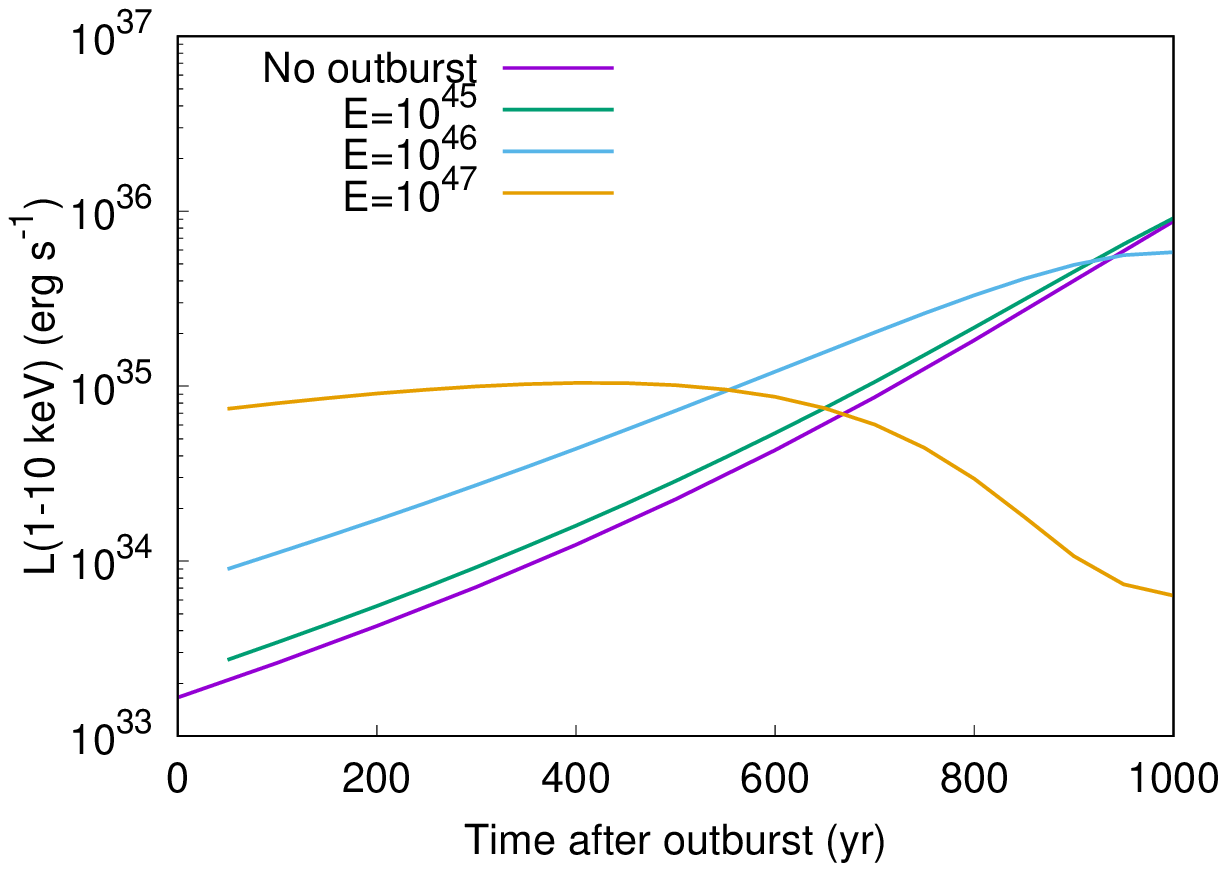}
\includegraphics[width=0.45\textwidth]{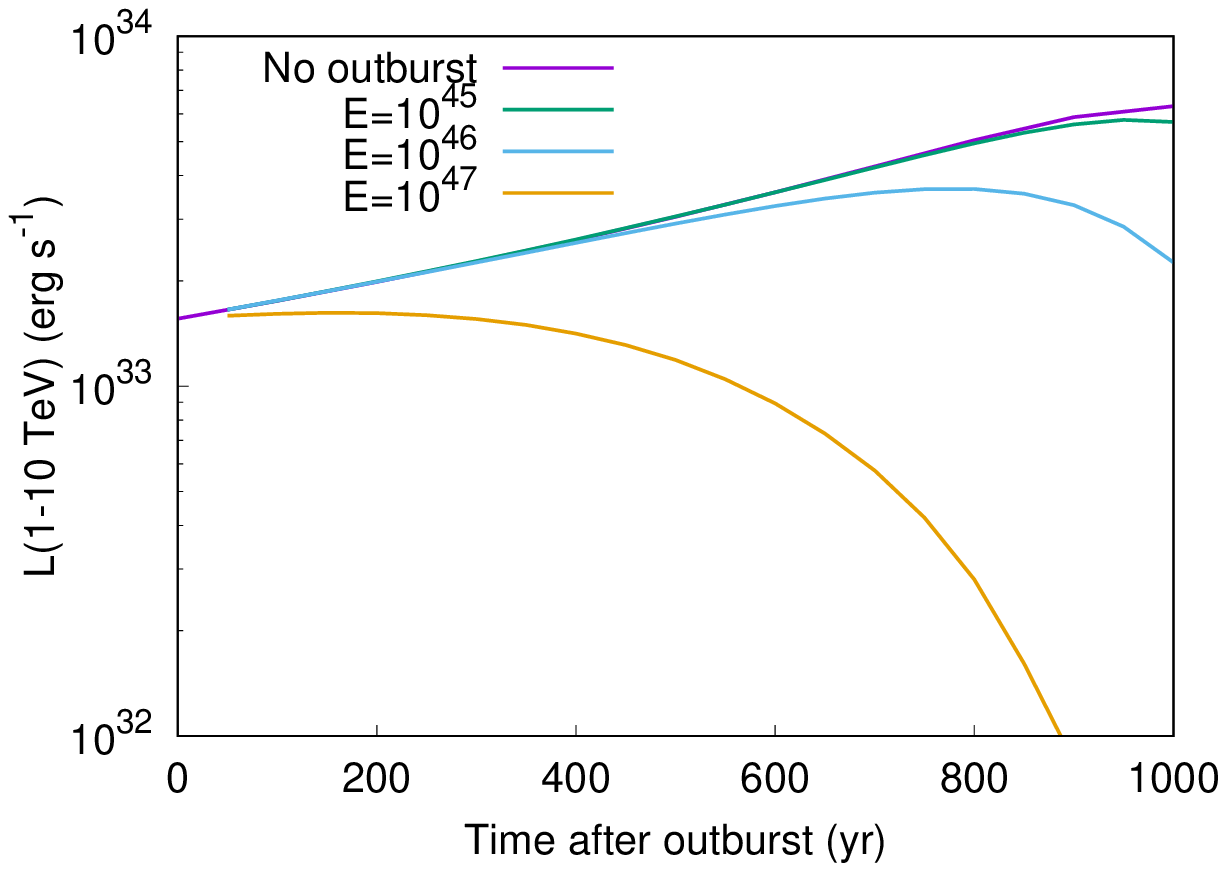}
\caption{Evolution of the luminosity in radio, X-rays and TeV (from top to bottom) when the injection is assumed to power the magnetic field. }
\label{fig:lummagevol}
\end{figure}

Table \ref{tab:lummag} shows the luminosities and efficiencies obtained if the energy released is injected into the magnetic field of the nebula.
In this case, we would expect a variation at radio and X-rays energies, since the enhancement of the magnetic field only affects the synchrotron radiation.
In fact, by increasing the efficiency of the particle losses in synchrotron, we would expect less radiation at higher energies. 
We observe this below.
In general, the increase in the luminosity is quite similar for all the values of $E_{out}$.
Note that this mechanism is slightly more efficient, because all the energy injected feeds the magnetic field, which then affects all relevant particles. 
Instead, in the case of a direct injection into particles, the energetics is spread into a wide population, and not all injected particles contribute to radio and X-rays.

\begin{figure*}
\centering
\includegraphics[width=0.45\textwidth]{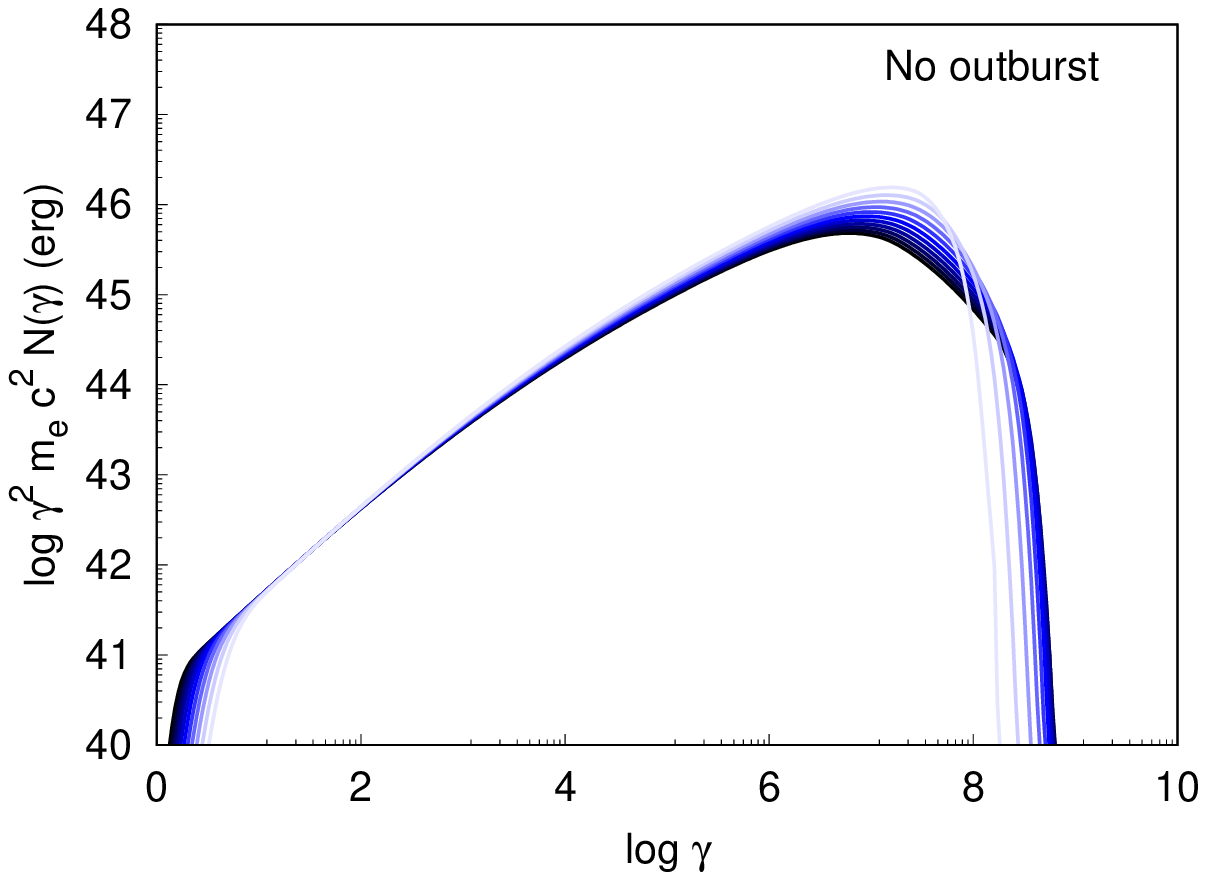}
\includegraphics[width=0.45\textwidth]{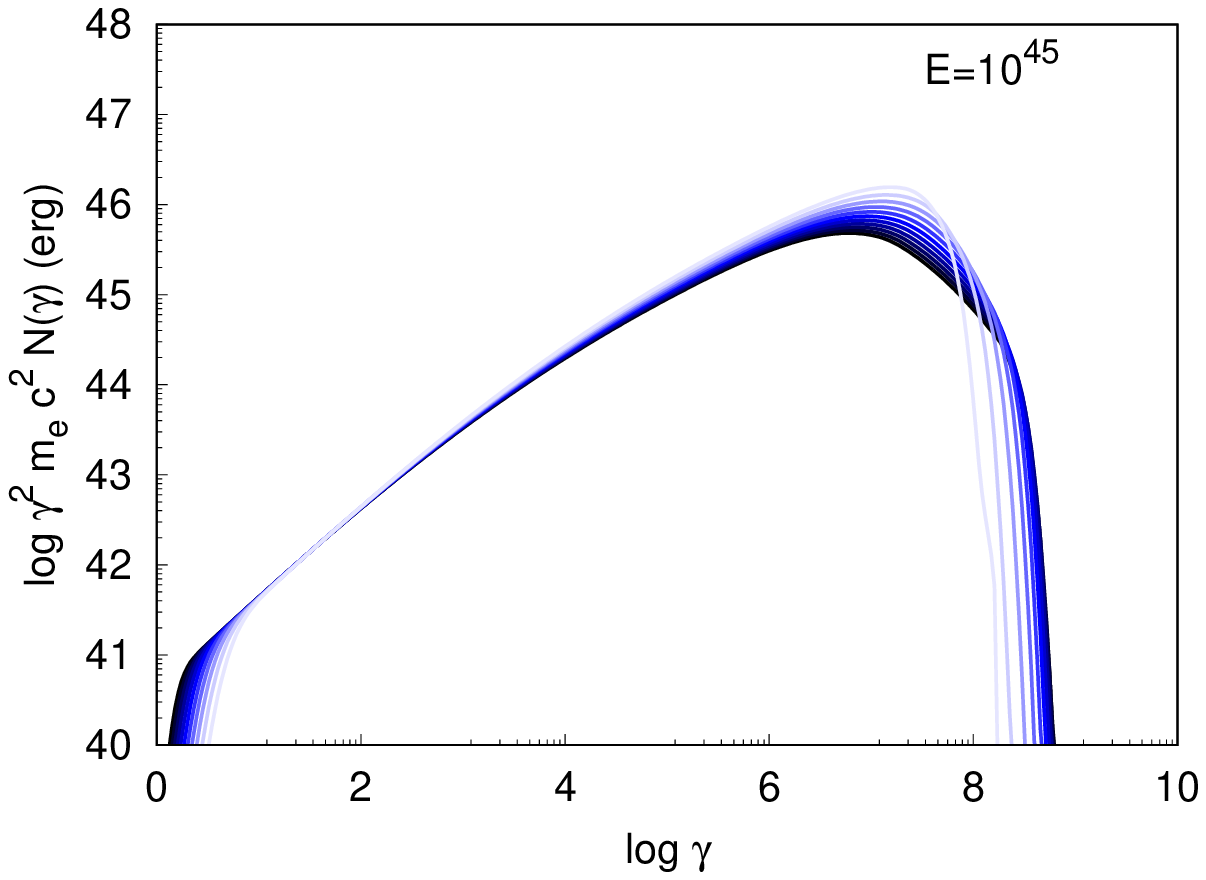}
\includegraphics[width=0.45\textwidth]{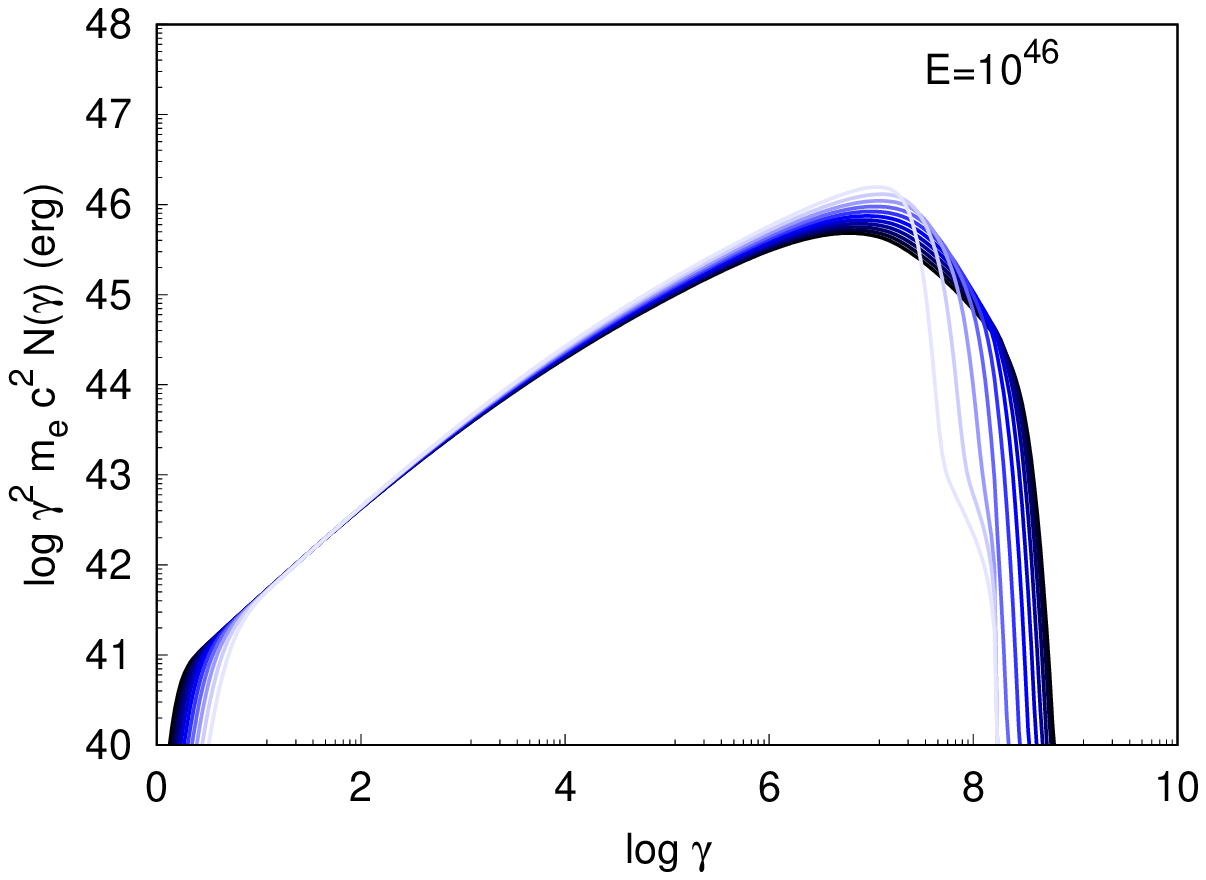}
\includegraphics[width=0.45\textwidth]{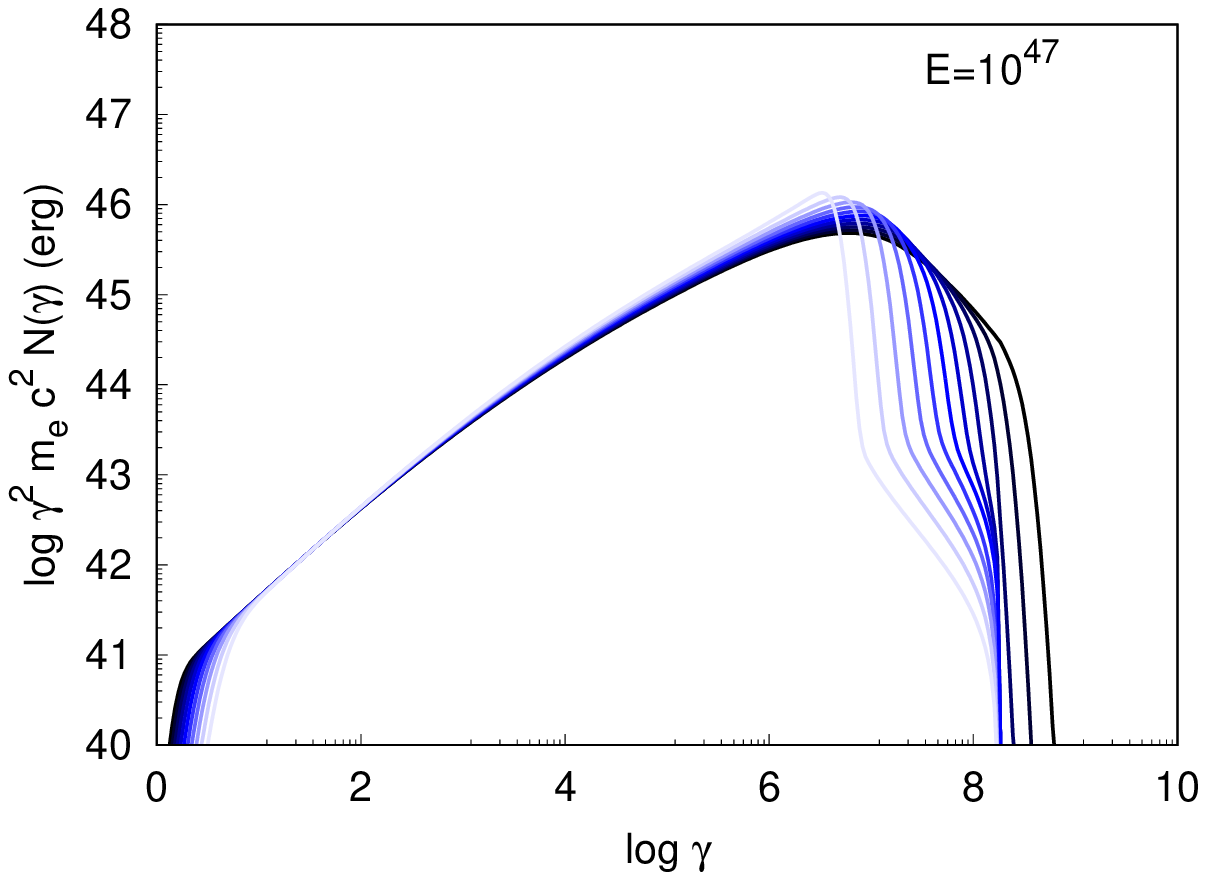}
\caption{Evolution of the pairs distribution function in time when the injection is assumed to power the magnetic field.}
\label{fig:elecmag}
\end{figure*}

Figure \ref{fig:lummagevol} shows the long term evolution of the radio, X-rays and VHE luminosities.
As in the previous case, the decay of the luminosity is very slow in comparison with what we see observationally, but its evolution in time is completely different.
For $E_{out}=10^{45}$ erg, the light curves are quite similar to the non-outburst case, because the energy injected is not enough to make a significant effect on the mean magnetic field. This changes for the other two cases ($E_{out}=10^{46}$  and $10^{47}$ erg).
For the latter, we observe that the X-rays luminosity decreases going below the non-outburst curves.
The same happens at VHE, with this happening even sooner.
The reason for that is found when we look at Figure \ref{fig:elecmag}, where we show the evolution of the pair population for each case.
The increase of the magnetic field modifies the particle distribution function, burning off the high energy particles which contribute specially to X-rays and VHE. The low energy particles spectrum (before the energy break) remains almost constant and the variation of the radio luminosity is explained only by the variation of the magnetic field.

Note that the scale in the x-axis of Figures 1 and 2 is long, and the luminosity remains high.
We note that the generic result here is that if the burst injects particles or powers the magnetic field into the pre-existing PWN, but there is no change the time scale for the losses, there is only an increase (rather than an increase and a decrease) of the luminosity, for periods as long as thousands of years.
The evolution in such scales may be affected by other dynamical processes, like reverberation, which in the case of 
J1834.9--0846 may be significant in the forthcoming hundreds of years (see \cite{Torres2017}).

\subsection{Burst powering of the magnetic field, augmented decay}
\label{secmag2}

We shall finally consider that the burst still powers the magnetic field, but that it decays in a time scale much shorter than that provided by adiabatic losses with the PWN expansion velocity.
This can be the result of a perturbation associated with the burst that travels at much higher velocities, or to other processes of damping.
For instance, in the starquake scenario \citep{blaes1989}, the cracking of the neutron star crust due to the presence of magnetic field tensions release energy and induce the propagation of an Alfv\'en wave in the outer magnetosphere.
In low density media, an Alfv\'en wave propagates approximately at the speed of light.
We shall assume that either this wave, or other sort of perturbation reaches the PWN, exerting work onto the ambient medium. 
This is, however, speculative. Whereas \cite{Thompson1998} found that an Alfv\'en wave could convert to so-called fast modes, and escape the magnetosphere, it is unclear whether they could reach so far distances as the termination shock \citep{Li-Zrake2019}.
Alfv\'en waves have also been noted as a possible origin for FRBs (see e.g., \cite{Lu2020}) although the location is thought to be closer to the pulsar.
It is not our intention here to enter into details regarding how such a wave could reach to the PWN, neither marry to the idea that such a scenario is in fact in place.
We rather focus on what would happen in this situation (i.e., a magnetic field perturbation travelling at the speed of light) at a phenomenological level.
This is motivated by the fact that a large increase of PWN fluxes in relatively short times scales have actually been observed. 

In practice, we consider that the evolution equation of the magnetic field energy of the perturbation $U'_B$ is
\begin{equation}
\frac{dU'_B}{dt}=\eta' L'(t)-\frac{U'_B}{R} \frac{dR}{dt}
\end{equation}
where $R$ is the radius of the perturbation wave which we take as $R \simeq c t$. After some algebra, the evolution of the perturbation yields
\begin{equation}
\frac{d(\Delta B)}{dt}=\frac{3 \eta'}{c^3} \frac{L'(t)}{\Delta B t^3}-\frac{2 \Delta B}{t},
\end{equation}
where $\Delta B$ is the magnetic field due to the additional injection (the total field being $B_{tot} = B_{pwn} + \Delta B$. Note that in this phenomenological approach, the additional field injected has a faster evolution in time than the originally residing in the nebula.
Regarding the energy injection, we assume that it has the form
\begin{equation}
L'(t)=L'_0 e^{-\frac{t-t_0}{t_{decay}}}
\end{equation}
where $t_0$ is the time when the injection starts, $t_{decay}$ the decay injection timescale and $L'_0$ the initial injection luminosity.
If the total energy injected is $E_{out}$, a fraction of such energy $\eta'$ will sustain the magnetic perturbation.
Thus, we can determine the value of $L'_0$ with the energy conservation
\begin{equation}
E_{out}=\int_{t_0}^\infty L'_0 e^{-\frac{t-t_0}{t_{decay}}} \mathrm{d}t
\end{equation}
resulting $L'_0=\eta' E_{out}/t_{decay}$.
For simplicity, we assume $\eta'=1$.
Thus, the final injection law yields
\begin{equation}
L'(t)=\frac{E_{out}}{t_{decay}} e^{-\frac{t-t_0}{t_{decay}}}
\end{equation}
which is reminiscent to equation 35 in \cite{blaes1989} for the Alfv\'en wave luminosity.

\begin{table*}
\scriptsize
\centering
\caption{Radio, X-ray and VHE luminosities and efficiencies obtained from 1 year simulations after the outburst. \label{tab:lumpar}}
\begin{tabular}{lcccccccc}
\hline
 & $E_{out}$ (erg) & $\delta$ & $L_{radio}$ ($10^{28}$ erg s$^{-1}$) & $L_{radio}/L_{sd}$ & $L_X$ ($10^{33}$ erg s$^{-1}$) & $L_X/L_{sd}$ & $L_{vhe}$ ($10^{33}$ erg s$^{-1}$) & $L_{vhe}/L_{sd}$\\
\hline
No outburst &  &  & 3.47 & $1.7 \cdot 10^{-6}$ & 1.67 & 0.08 & 1.56 & 0.074\\
\hline
\multirow{9}{*}{$E_{max}=1$ TeV} & \multirow{3}{*}{$10^{45}$} & 2.5 & 3.52 & $1.7 \cdot 10^{-6}$ & 1.67 & 0.08 & 1.56 & 0.074\\
 & & 2 & 4.28 & $2.0 \cdot 10^{-6}$ & 1.67 & 0.08 & 1.56 & 0.074\\
 & & 1.5 & 4.05 & $1.9 \cdot 10^{-6}$ & 1.67 & 0.08 & 1.56 & 0.074\\
\cline{2-9}
 & \multirow{3}{*}{$10^{46}$} & 2.5 & 3.99 & $1.9 \cdot 10^{-6}$ & 1.67 & 0.08 & 1.56 & 0.074\\
 & & 2 & 11.6 & $5.5 \cdot 10^{-6}$ & 1.67 & 0.08 & 1.56 & 0.074\\
 & & 1.5 & 9.22 & $4.4 \cdot 10^{-6}$ & 1.67 & 0.08 & 1.57 & 0.075\\
\cline{2-9}
 & \multirow{3}{*}{$10^{47}$} & 2.5 & 8.62 & $4.1 \cdot 10^{-6}$ & 1.67 & 0.08 & 1.56 & 0.074\\
 & & 2 & 84.9 & $4.0 \cdot 10^{-5}$ & 1.67 & 0.08 & 1.56 & 0.074\\
 & & 1.5 & 61.0 & $2.9 \cdot 10^{-5}$ & 1.67 & 0.08 & 1.62 & 0.077\\
\hline
\multirow{9}{*}{$E_{max}=30$ TeV} & \multirow{3}{*}{$10^{45}$} & 2.5 & 3.52 & $1.7 \cdot 10^{-6}$ & 1.67 & 0.08 & 1.56 & 0.074\\
 & & 2 & 4.13 & $2.0 \cdot 10^{-6}$ & 1.69 & 0.08 & 1.58 & 0.075\\
 & & 1.5 & 3.58 & $1.7 \cdot 10^{-6}$ & 1.94 & 0.092 & 1.73 & 0.082\\
\cline{2-9}
 & \multirow{3}{*}{$10^{46}$} & 2.5 & 3.99 & $1.9 \cdot 10^{-6}$ & 1.67 & 0.08 & 1.56 & 0.074\\
 & & 2 & 10.1 & $4.8 \cdot 10^{-6}$ & 1.85 & 0.088 & 1.77 & 0.084\\
 & & 1.5 & 4.54 & $2.2 \cdot 10^{-6}$ & 4.42 & 0.21 & 3.21 & 0.15\\
\cline{2-9}
 & \multirow{3}{*}{$10^{47}$} & 2.5 & 8.65 & $4.1 \cdot 10^{-6}$ & 1.67 & 0.08 & 1.56 & 0.074\\
 & & 2 & 69.5 & $3.3 \cdot 10^{-5}$ & 3.52 & 0.17 & 3.65 & 0.17\\
 & & 1.5 & 14.2 & $6.8 \cdot 10^{-6}$ & 29.2 & 1.39 & 18.1 & 0.86\\
\hline
\end{tabular}
\end{table*}

\begin{table*}
\scriptsize
\centering
\caption{Radio, X-ray and VHE luminosities and efficiencies obtained from the 1 year simulations after the outburst in the case the energy is injected into the magnetic field. \label{tab:lummag}}
\begin{tabular}{lccccccc}
\hline
$E_{out}$ (erg) & $B$ ($\mu$G) & $L_{radio}$ ($10^{28}$ erg s$^{-1}$) & $L_{radio}/L_{sd}$ & $L_X$ ($10^{33}$ erg s$^{-1}$) & $L_X/L_{sd}$ & $L_{vhe}$ ($10^{33}$ erg s$^{-1}$) & $L_{vhe}/L_{sd}$\\
\hline
No outburst & 4.4 & 3.47 & $1.7 \cdot 10^{-6}$ & 1.67 & 0.08 & 1.56 & 0.074\\
$10^{45}$ & 4.9 & 3.94 & $1.9 \cdot 10^{-6}$ & 2.18 & 0.10 & 1.56 & 0.074\\
$10^{46}$ & 8.1 & 7.10 & $3.4 \cdot 10^{-6}$ & 7.37 & 0.35 & 1.56 & 0.074\\
$10^{47}$ & 22.0 & 22.4 & $1.1 \cdot 10^{-5}$ & 68.7 & 3.3 & 1.56 & 0.074\\
\hline
\end{tabular}
\end{table*}

\begin{figure*}
\centering
\includegraphics[width=0.45\textwidth]{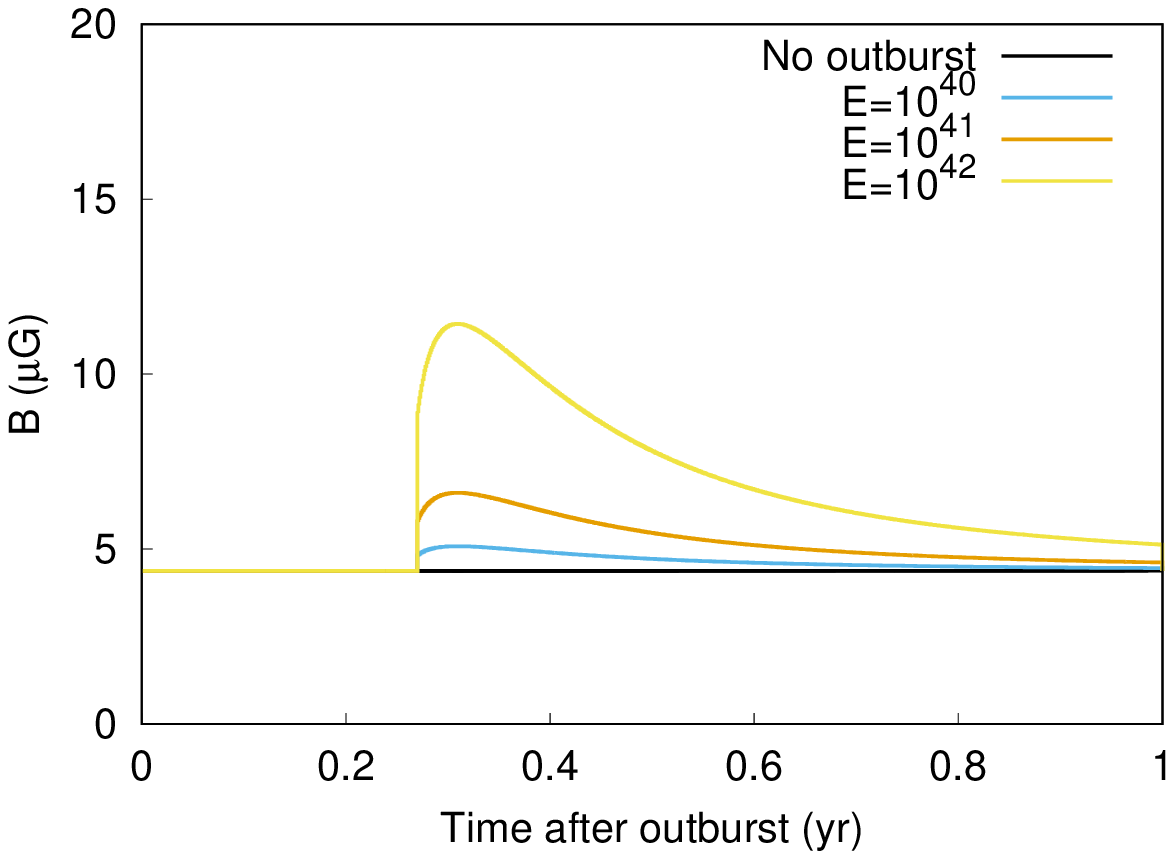}
\includegraphics[width=0.45\textwidth]{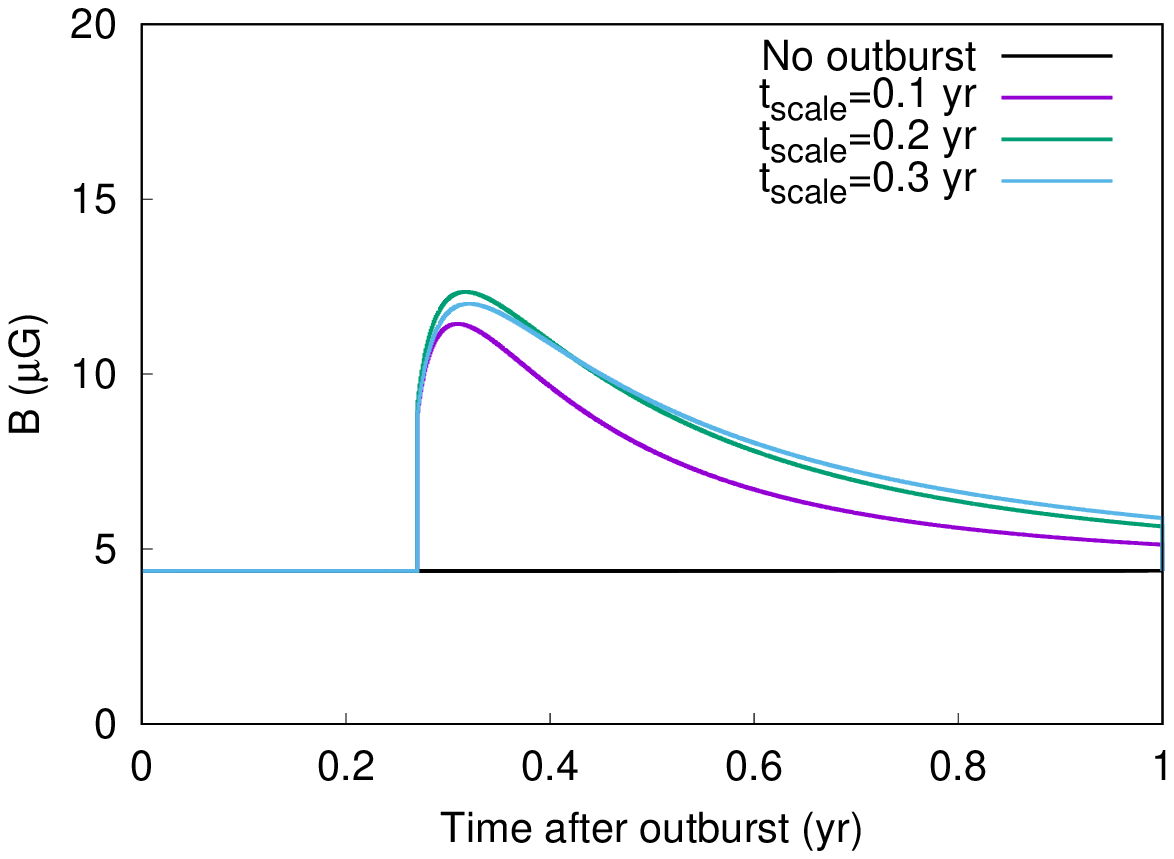}
\caption{Left: Evolution of the magnetic field when $t_{decay}=0.1$ yr, other parameters are as in Table \ref{tab:lummag2}. 
Right: Evolution of the magnetic field varying $t_{decay}$. In this case, we fix $\eta'=1$ and $E_{out}=10^{42}$ erg. }
\label{fig:bevol}
\end{figure*}

\begin{figure*}
\centering
\includegraphics[width=0.45\textwidth]{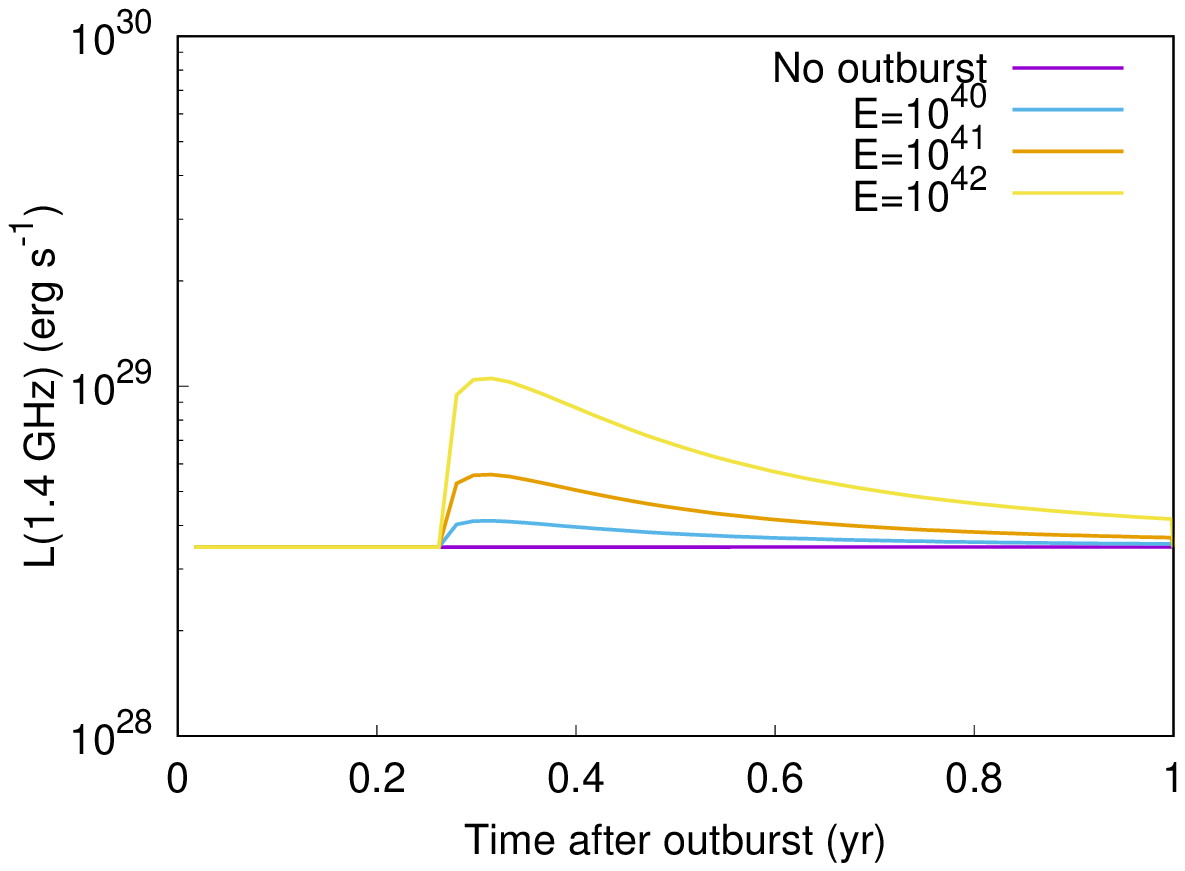}
\includegraphics[width=0.45\textwidth]{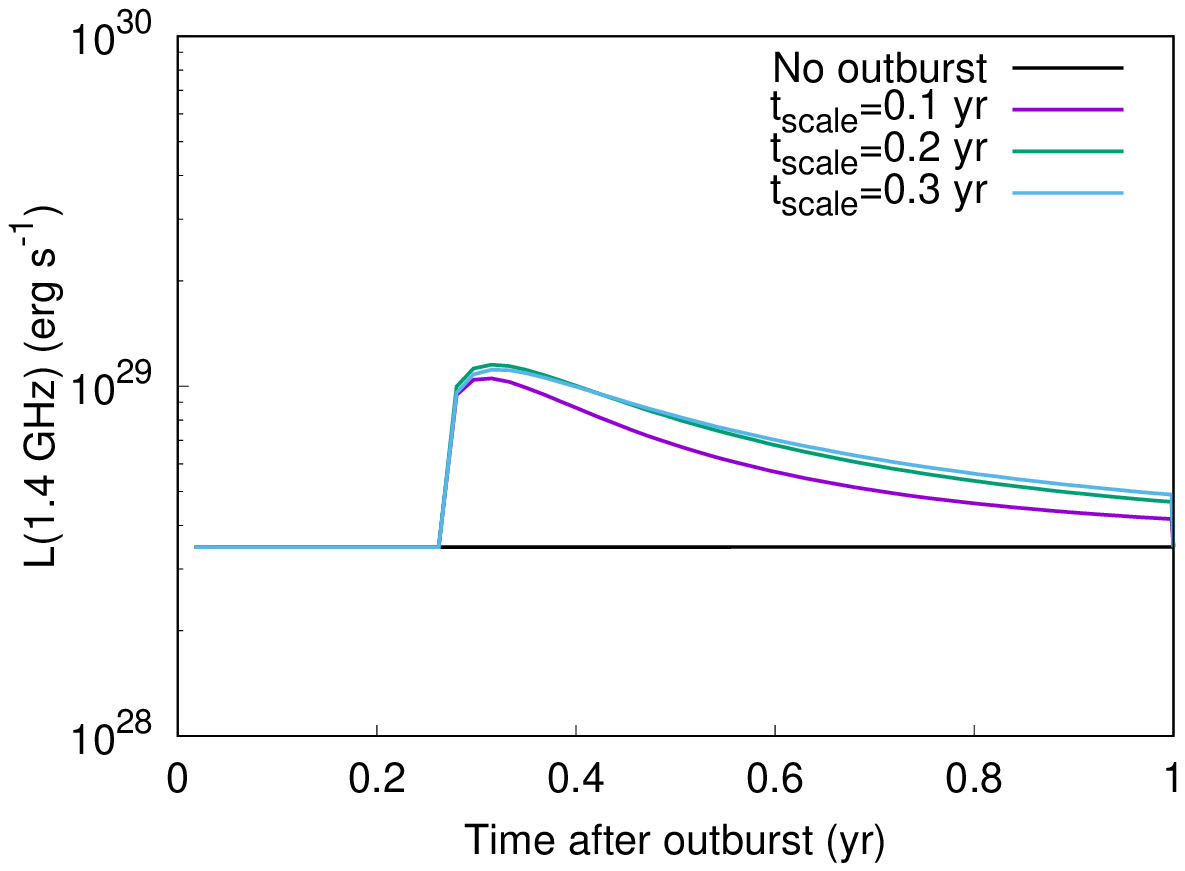}
\includegraphics[width=0.45\textwidth]{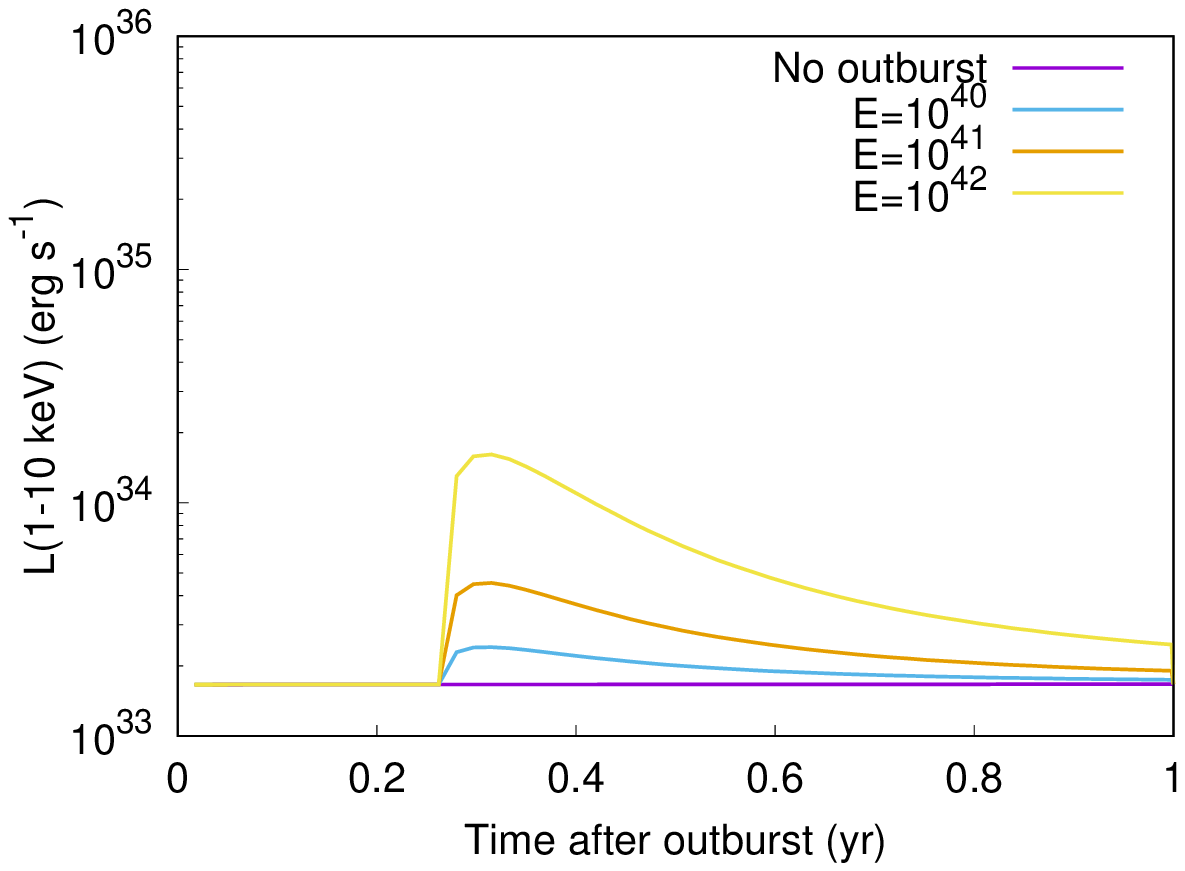}
\includegraphics[width=0.45\textwidth]{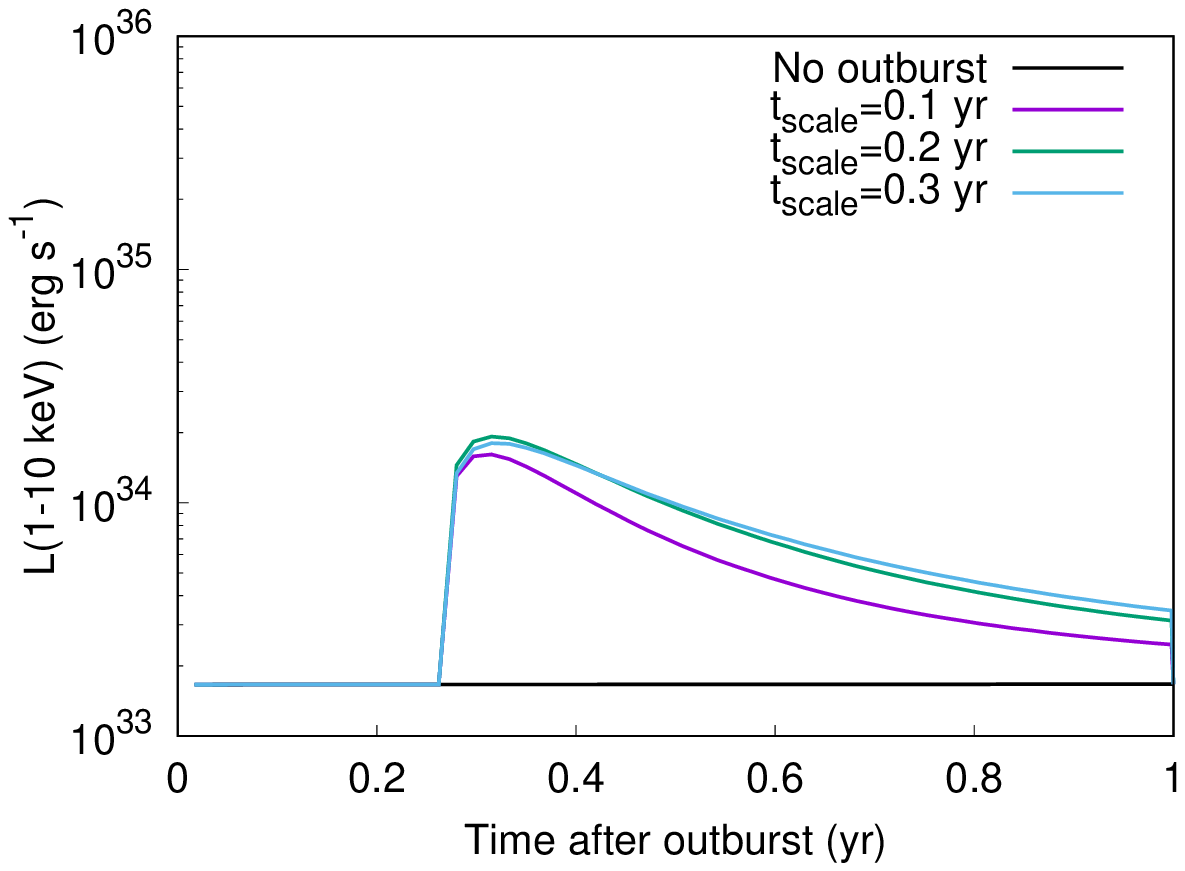}
\includegraphics[width=0.45\textwidth]{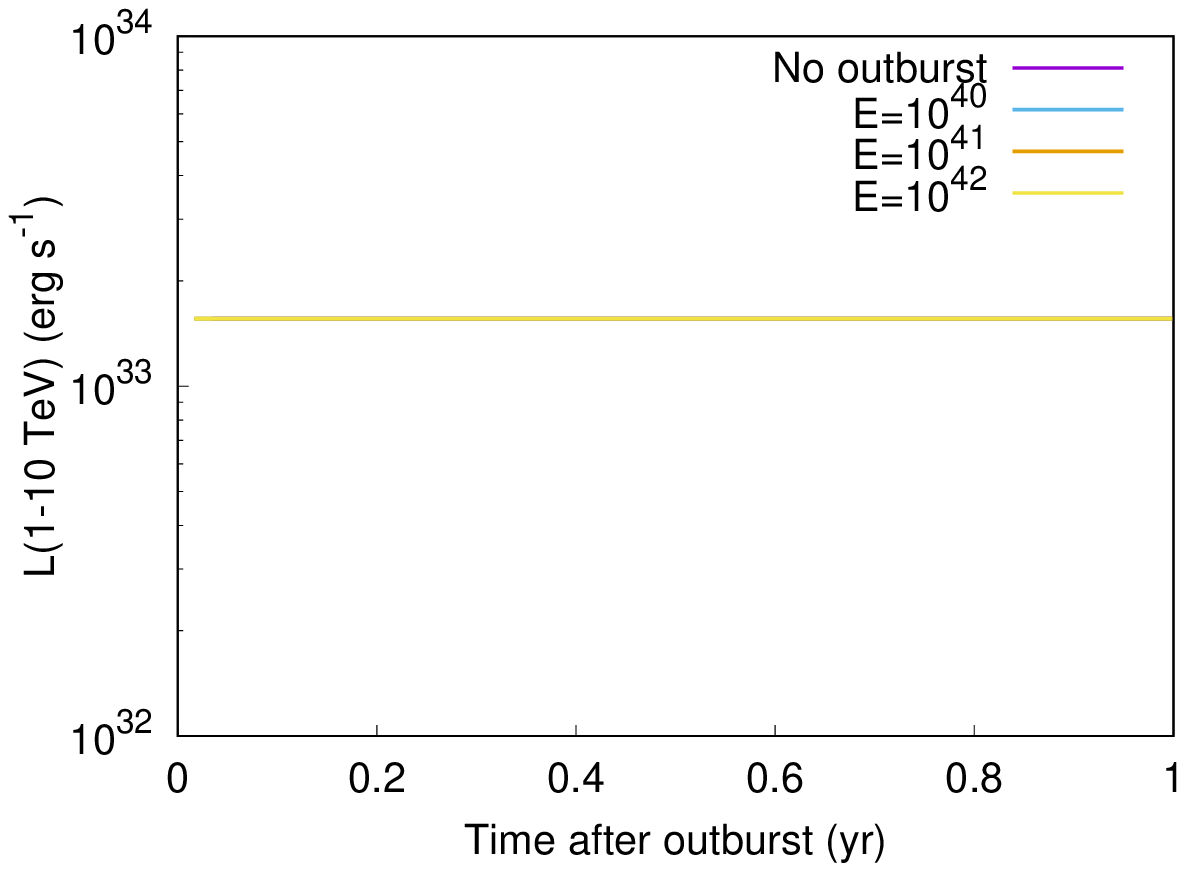}
\includegraphics[width=0.45\textwidth]{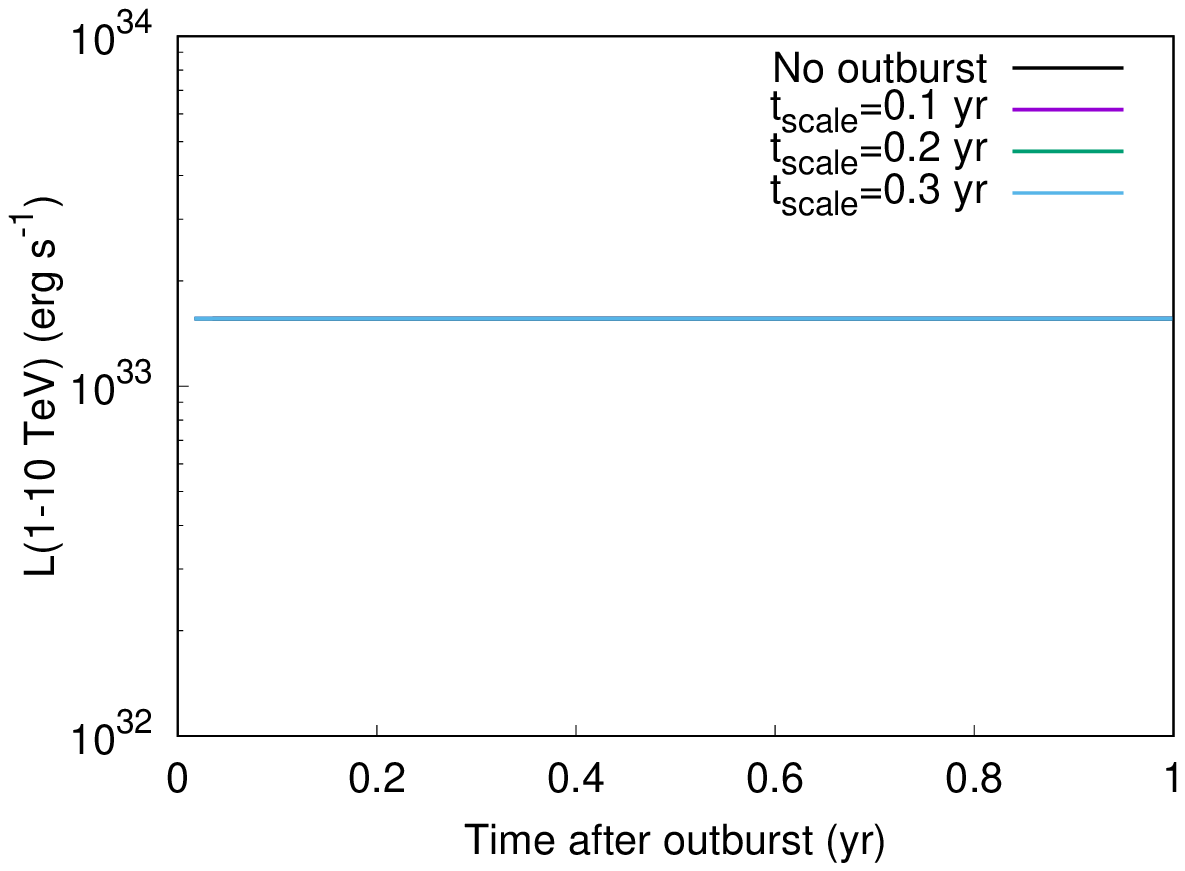}
\caption{Evolution of the luminosity in radio, X-rays and TeV when $t_{decay}=0.1$ yr.}
\label{fig:blumevol}
\end{figure*}

Table \ref{tab:lummag2} summarizes the results.
Note that the injected energy considered in each case ($E_{out}=10^{40}$, $10^{41}$ and $10^{42}$ erg) is much lower than in the previous simulations.
The efficiency of this mechanism is larger, the X-ray flux increases one order of magnitude already with $E_{out}=10^{42}$ erg.
The radius of the volume containing the extra magnetic energy expands at the speed of light exerting force into the medium.
The perturbation affects the PWN when it reaches the termination shock.
The evolution of the magnetic field is shown in Figure \ref{fig:bevol}.
The jump in Figure~\ref{fig:bevol} (and also in Figure~\ref{fig:blumevol}) is produced by the time needed by the perturbation to reach the PWN.
The parameter $t_{decay}$ spreads the energy injection in a larger or smaller timescale.
The right panel of Figure \ref{fig:bevol} shows how the magnetic field changes by varying this parameter.
The larger is $t_{decay}$, the softer is the magnetic field decay to quiescence.
As we can see in Figure \ref{fig:blumevol}, the decay timescales are shorter.
The luminosity at VHE remains unaffected by the increase of the magnetic field, while we observe a similar effect in radio and X-rays.
\begin{table*}
\scriptsize
\centering
\caption{X-ray and VHE luminosities and efficiencies obtained from the 1 year simulations after the outburst in the case of a magnetic field pertubation travelling at the speed of light. \label{tab:lummag2}}
\begin{tabular}{lccccccc}
\hline
$E_{out}$ (erg) & $B_{max}$ ($\mu$G) & $L_{radio}$ ($10^{28}$ erg s$^{-1}$) & $L_{radio}/L_{sd}$ & $L_X$ ($10^{33}$ erg s$^{-1}$) & $L_X/L_{sd}$ & $L_{vhe}$ ($10^{33}$ erg s$^{-1}$) & $L_{vhe}/L_{sd}$\\
\hline
No outburst & 4.4 & 3.47 & $1.7 \cdot 10^{-6}$ & 1.67 & 0.08 & 1.56 & 0.074\\
$10^{40}$ & 5.3 & 4.28 & $2.0 \cdot 10^{-6}$ & 2.60 & 0.12 & 1.56 & 0.074\\
$10^{41}$ & 7.1 & 6.10 & $2.9 \cdot 10^{-6}$ & 5.41 & 0.26 & 1.56 & 0.074\\
$10^{42}$ & 13.1 & 12.3 & $5.9 \cdot 10^{-6}$ & 21.7 & 1.03 & 1.56 & 0.074\\
\hline
\end{tabular}
\end{table*}

We do not show the effects of lower energy flares (i.e., lower total energetics than the ones used in Figure 5), but one can gather immediately that the emitted flux is increasingly lower. 
Also, we note that it is very unlikely that GBM or Swift missed bright burst along their lifetime. 
However, we recall that as said above, the bursts from J1119 we are taken as examples  emitted a total energy of $3.7 \times 10^{38}$ and $5.2 \times 10^{38}$ erg just between 8 and 200 keV, and lasted 36 and 186 ms, respectively.
Whereas in comparison with the energetics we use in Figure 5, where we show temporal variability in the PWN happening within 1 year, these burst energetics in X-rays are much lower, we are not constrained by the X-ray measurements: Our total energetics --computed using Eqs. 17 and 18 --must be invested  into producing the X-ray burst, but also into local particles and magnetic field powering for which GBM or Swift are blind.  That is, the X-ray measurement should function as a lower limit to the total power available.

\section{A connection with FRBs}

Fast Radio Burst (FRBs) are pulses at radio frequencies with very short durations (some milliseconds) (see \citealt{Lorimer2007} and subsequent works) with large dispersion measures (DMs).
The burst mechanism is still under debate, but many studies propose that the emission could be powered by the decaying magnetic field of a  magnetar, similarly to what we are considering in sections \ref{secmag} and \ref{secmag2} \citep{Popov2013,Lyubarsky2014,Katz2016,Lu2016,Murase2016,Beloborodov2017,Kumar2017,Metzger2017,Nicholl2017,Margalit2018,Yang2019}.
The active repeating FRBs are believed to be produced by young magnetars usually hundreds of years old, whereas older magnetars similar to the observed Galactic magnetars may also produce less energetic FRBs with a lower repetition rate.

Recently, the Canadian Hydrogen Intensity Mapping Experiment (CHIME)  \citep{CHIME} and STARE-2 \citep{STARE2} discovered FRB~200428, an FRB-like event with an energy $\sim 3 \times 10^{34}$ erg, which is about three orders of magnitude beyond the radio emission detected from magnetars but is $\sim$ 20 times lower than the energy of faintest cosmological FRBs, that was in association with one hard X-ray burst \citep{HXMT-FRB,Mereghetti2020} from the Galactic magnetar, SGR 1935+2154.
The lightcurve of the concurrent non-thermal X-ray burst as detected by Insight-HMXT in the 1-250 keV energy band \citep{HXMT-FRB} showed two hard peaks with a separation of $\sim 30$ ms, consistent with the separation between the two bursts in FRB~200428. 
This suggests that Galactic old magnetars can make FRBs, even though does not appear to be common \citep{FAST-FRB}.

A difference between an old and a nascent FRB is that the ambient medium is much denser in the latter case, whereas for an older PWN the medium is optically thin and we do not need to account for synchrotron self-absorption (SSA).
On the contrary, interactions between FRB-related flares and the associated PWN should be more common and efficient for young magnetars. The effect of synchrotron heating in a PWN by repeated FRBs has been studied by \cite{Yang2016} and \cite{LiQiao-Chu2020}.
There, synchrotron self-absorption could be important under certain conditions and the absorbed energy can be used to heat up the nebula.
Even though FRB emission likely comes from a magnetar magnetosphere \citep{Kumar2017,YangZhang2018,Lu2020,Luo2020,Zhang2020}, models invoking synchrotron maser in a relativistic shock (which can overcome the induced Compton scattering constraint) have been also discussed in the literature \citep{Lyubarsky2014,Sironi2019,Metzger2019,Beloborodov2020}. In both cases, the energy reservoir is likely the magnetar $B$ field energy \citep{Margalit2018}, and high energy emission concurrent with radio emission is predicted (with the latter type of model typically predict a higher luminosity \citep{Zhang2020}. In any case, the emission is difficult to detect from a cosmological distance.

Further observing PWN J1119--6127 is critical to understand the long-term behavior of its PWN and the return to its quiescent state, if such occurs.
In case that the increase of the X-ray flux in PWN J1119--6127 is confirmed, and a decay follows in a short timescales, there could be a relation between the mechanism that enhanced such luminosity and FRBs.
On the other hand, if the luminosity in PWN J1119--6127 stays high and constant, it would be consistent with injection of energy either in particles or in magnetic field
under the prior conditions of the nebula.

\section{Conclusions}

We have seen that if a powerful magnetar burst injects energy into a relatively evolved PWN (of few thousands years); either into the particle population or into the magnetic field, there might be an increase of luminosities. 
Depending on the specific injection and the magnetic field of the nebula the increase can happen in radio only, or in radio and higher energies.
However, in both cases, if the new injected particles or the new field are subject to the losses in an equal way than in the pre-burst nebula, 
the time scale for the decay of the enhanced radiation is thousands of years. 
This may not be used to explain short time scales variabilities as observed happening in the PWN of magnetar J1119--6127, as described in the introduction.

A fast expanding perturbation of the magnetic field followed by an augmented decay seems more plausible in order to get a luminosity evolution of the same magnitude in variability and with shorter time scales.
On the one hand, the energetics is eased (a less energetic burst more efficiently produces a larger increase in luminosity). On the other hand, the time scale can be governed by a damping process different from the one operating in the pre-burst PWN.
Such mechanism could also explain why there are so few PWNe detected in magnetars:
A sudden strong increase of the magnetic field following a magnetar burst could burn off all the high energy particles residing inside its PWN, which would then become almost undetectable until it gets filled again with new particles accelerated from the spin-down energy loss of the magnetar.
Given that magnetars are generally of low spin-down power, the latter can take a significant time, and never produce a bright PWN again.


\bibliographystyle{mnras} 

\section*{Acknowledgements}
%
This work has been supported by grants PGC2018-095512-B-I00, SGR2017-1383, and AYA2017-92402-EXP.

\bibliography{pwn}

\begin{thebibliography}{}
\makeatletter
\relax
\def\mn@urlcharsother{\let\do\@makeother \do\$\do\&\do\#\do\^\do\_\do\%\do\~}
\def\mn@doi{\begingroup\mn@urlcharsother \@ifnextchar [ {\mn@doi@}
  {\mn@doi@[]}}
\def\mn@doi@[#1]#2{\def\@tempa{#1}\ifx\@tempa\@empty \href
  {http://dx.doi.org/#2} {doi:#2}\else \href {http://dx.doi.org/#2} {#1}\fi
  \endgroup}
\def\mn@eprint#1#2{\mn@eprint@#1:#2::\@nil}
\def\mn@eprint@arXiv#1{\href {http://arxiv.org/abs/#1} {{\tt arXiv:#1}}}
\def\mn@eprint@dblp#1{\href {http://dblp.uni-trier.de/rec/bibtex/#1.xml}
  {dblp:#1}}
\def\mn@eprint@#1:#2:#3:#4\@nil{\def\@tempa {#1}\def\@tempb {#2}\def\@tempc
  {#3}\ifx \@tempc \@empty \let \@tempc \@tempb \let \@tempb \@tempa \fi \ifx
  \@tempb \@empty \def\@tempb {arXiv}\fi \@ifundefined
  {mn@eprint@\@tempb}{\@tempb:\@tempc}{\expandafter \expandafter \csname
  mn@eprint@\@tempb\endcsname \expandafter{\@tempc}}}

\bibitem[\protect\citeauthoryear{{Aleksi{\'c}} et~al.,}{{Aleksi{\'c}}
  et~al.}{2013}]{Aleksic2013b}
{Aleksi{\'c}} J.,  et~al., 2013, \mn@doi [\aap] {10.1051/0004-6361/201220275},
  \href {https://ui.adsabs.harvard.edu/abs/2013A&A...549A..23A} {549, A23}

\bibitem[\protect\citeauthoryear{{Anderson} et~al.,}{{Anderson}
  et~al.}{2012}]{Anderson2012}
{Anderson} G.~E.,  et~al., 2012, \mn@doi [\apj] {10.1088/0004-637X/751/1/53},
  \href {https://ui.adsabs.harvard.edu/abs/2012ApJ...751...53A} {751, 53}

\bibitem[\protect\citeauthoryear{{Bandiera}, {Bucciantini}, {Martin}, {Olmi}
  \& {Torres}}{{Bandiera} et~al.}{2020}]{Martin2020}
{Bandiera} R.,  {Bucciantini} N.,  {Martin} J.,  {Olmi} B.,   {Torres} D.~F.,
  2020, submitted

\bibitem[\protect\citeauthoryear{{Beloborodov}}{{Beloborodov}}{2017}]{Beloborodov2017}
{Beloborodov} A.~M.,  2017, \mn@doi [\apjl] {10.3847/2041-8213/aa78f3}, \href
  {https://ui.adsabs.harvard.edu/abs/2017ApJ...843L..26B} {843, L26}

\bibitem[\protect\citeauthoryear{{Beloborodov}, {Lundman}  \&
  {Levin}}{{Beloborodov} et~al.}{2020}]{Beloborodov2020}
{Beloborodov} A.~M.,  {Lundman} C.,   {Levin} Y.,  2020, \mn@doi [\apj]
  {10.3847/1538-4357/ab86a0}, \href
  {https://ui.adsabs.harvard.edu/abs/2020ApJ...897..141B} {897, 141}

\bibitem[\protect\citeauthoryear{{Beniamini}, {Hotokezaka}, {van der Horst}  \&
  {Kouveliotou}}{{Beniamini} et~al.}{2019}]{Beniamini2019}
{Beniamini} P.,  {Hotokezaka} K.,  {van der Horst} A.,   {Kouveliotou} C.,
  2019, \mn@doi [\mnras] {10.1093/mnras/stz1391}, \href
  {https://ui.adsabs.harvard.edu/abs/2019MNRAS.487.1426B} {487, 1426}

\bibitem[\protect\citeauthoryear{{Blaes}, {Blandford}, {Goldreich}  \&
  {Madau}}{{Blaes} et~al.}{1989}]{blaes1989}
{Blaes} O.,  {Blandford} R.,  {Goldreich} P.,   {Madau} P.,  1989, \mn@doi
  [\apj] {10.1086/167754}, \href
  {https://ui.adsabs.harvard.edu/abs/1989ApJ...343..839B} {343, 839}

\bibitem[\protect\citeauthoryear{{Blumer}, {Safi-Harb}  \&
  {McLaughlin}}{{Blumer} et~al.}{2017}]{blumer2017}
{Blumer} H.,  {Safi-Harb} S.,   {McLaughlin} M.~A.,  2017, \mn@doi [\apjl]
  {10.3847/2041-8213/aa9844}, \href
  {https://ui.adsabs.harvard.edu/abs/2017ApJ...850L..18B} {850, L18}

\bibitem[\protect\citeauthoryear{{Bochenek}, {Ravi}, {Belov}, {Hallinan},
  {Kocz}, {Kulkarni}  \& {McKenna}}{{Bochenek} et~al.}{2020}]{STARE2}
{Bochenek} C.~D.,  {Ravi} V.,  {Belov} K.~V.,  {Hallinan} G.,  {Kocz} J.,
  {Kulkarni} S.~R.,   {McKenna} D.~L.,  2020, arXiv e-prints, \href
  {https://ui.adsabs.harvard.edu/abs/2020arXiv200510828B} {p. arXiv:2005.10828}

\bibitem[\protect\citeauthoryear{{Cameron} et~al.,}{{Cameron}
  et~al.}{2005}]{Cameron2005}
{Cameron} P.~B.,  et~al., 2005, \mn@doi [\nat] {10.1038/nature03605}, \href
  {https://ui.adsabs.harvard.edu/abs/2005Natur.434.1112C} {434, 1112}

\bibitem[\protect\citeauthoryear{{Camilo}, {Ransom}, {Halpern}, {Reynolds},
  {Helfand}, {Zimmerman}  \& {Sarkissian}}{{Camilo} et~al.}{2006}]{Camilo2006}
{Camilo} F.,  {Ransom} S.~M.,  {Halpern} J.~P.,  {Reynolds} J.,  {Helfand}
  D.~J.,  {Zimmerman} N.,   {Sarkissian} J.,  2006, \mn@doi [\nat]
  {10.1038/nature04986}, \href
  {https://ui.adsabs.harvard.edu/abs/2006Natur.442..892C} {442, 892}

\bibitem[\protect\citeauthoryear{{Camilo}, {Ransom}, {Halpern}  \&
  {Reynolds}}{{Camilo} et~al.}{2007}]{Camilo2007}
{Camilo} F.,  {Ransom} S.~M.,  {Halpern} J.~P.,   {Reynolds} J.,  2007, \mn@doi
  [\apjl] {10.1086/521826}, \href
  {https://ui.adsabs.harvard.edu/abs/2007ApJ...666L..93C} {666, L93}

\bibitem[\protect\citeauthoryear{{Coti Zelati}, {Rea}, {Pons}, {Campana}  \&
  {Esposito}}{{Coti Zelati} et~al.}{2018}]{cotizelati2018}
{Coti Zelati} F.,  {Rea} N.,  {Pons} J.~A.,  {Campana} S.,   {Esposito} P.,
  2018, \mn@doi [\mnras] {10.1093/mnras/stx2679}, \href
  {https://ui.adsabs.harvard.edu/abs/2018MNRAS.474..961C} {474, 961}

\bibitem[\protect\citeauthoryear{{Frail}, {Kulkarni}  \& {Bloom}}{{Frail}
  et~al.}{1999}]{Frail1999}
{Frail} D.~A.,  {Kulkarni} S.~R.,   {Bloom} J.~S.,  1999, \mn@doi [\nat]
  {10.1038/18163}, \href
  {https://ui.adsabs.harvard.edu/abs/1999Natur.398..127F} {398, 127}

\bibitem[\protect\citeauthoryear{{Gaensler} et~al.,}{{Gaensler}
  et~al.}{2005}]{Gaensler2005}
{Gaensler} B.~M.,  et~al., 2005, \mn@doi [\nat] {10.1038/nature03498}, \href
  {https://ui.adsabs.harvard.edu/abs/2005Natur.434.1104G} {434, 1104}

\bibitem[\protect\citeauthoryear{{Gelfand}, {Slane}  \& {Zhang}}{{Gelfand}
  et~al.}{2009}]{Gelfand2009}
{Gelfand} J.~D.,  {Slane} P.~O.,   {Zhang} W.,  2009, \mn@doi [\apj]
  {10.1088/0004-637X/703/2/2051}, \href
  {http://adsabs.harvard.edu/abs/2009ApJ...703.2051G} {703, 2051}

\bibitem[\protect\citeauthoryear{{Goldreich} \& {Reisenegger}}{{Goldreich} \&
  {Reisenegger}}{1992}]{Goldreich1992}
{Goldreich} P.,  {Reisenegger} A.,  1992, \mn@doi [\apj] {10.1086/171646},
  \href {https://ui.adsabs.harvard.edu/abs/1992ApJ...395..250G} {395, 250}

\bibitem[\protect\citeauthoryear{{Gonzalez} \& {Safi-Harb}}{{Gonzalez} \&
  {Safi-Harb}}{2003}]{gonzalez2003}
{Gonzalez} M.,  {Safi-Harb} S.,  2003, \mn@doi [\apjl] {10.1086/377070}, \href
  {https://ui.adsabs.harvard.edu/abs/2003ApJ...591L.143G} {591, L143}

\bibitem[\protect\citeauthoryear{{G{\"o}{\u{g}}{\"u}{\textcommabelow s}}
  et~al.,}{{G{\"o}{\u{g}}{\"u}{\textcommabelow s}} et~al.}{2016}]{gogus2016}
{G{\"o}{\u{g}}{\"u}{\textcommabelow s}} E.,  et~al., 2016, \mn@doi [\apjl]
  {10.3847/2041-8205/829/2/L25}, \href
  {https://ui.adsabs.harvard.edu/abs/2016ApJ...829L..25G} {829, L25}

\bibitem[\protect\citeauthoryear{{Granot}, {Gill}, {Younes}, {Gelfand},
  {Harding}, {Kouveliotou}  \& {Baring}}{{Granot} et~al.}{2017}]{Granot2017}
{Granot} J.,  {Gill} R.,  {Younes} G.,  {Gelfand} J.,  {Harding} A.,
  {Kouveliotou} C.,   {Baring} M.~G.,  2017, \mn@doi [\mnras]
  {10.1093/mnras/stw2554}, \href
  {http://adsabs.harvard.edu/abs/2017MNRAS.464.4895G} {464, 4895}

\bibitem[\protect\citeauthoryear{{Harding}, {Contopoulos}  \&
  {Kazanas}}{{Harding} et~al.}{1999}]{Harding1999}
{Harding} A.~K.,  {Contopoulos} I.,   {Kazanas} D.,  1999, \mn@doi [\apjl]
  {10.1086/312339}, \href
  {https://ui.adsabs.harvard.edu/abs/1999ApJ...525L.125H} {525, L125}

\bibitem[\protect\citeauthoryear{{Katz}}{{Katz}}{2016}]{Katz2016}
{Katz} J.~I.,  2016, \mn@doi [\apj] {10.3847/0004-637X/826/2/226}, \href
  {https://ui.adsabs.harvard.edu/abs/2016ApJ...826..226K} {826, 226}

\bibitem[\protect\citeauthoryear{{Kennea}, {Lien}, {Marshall}, {Palmer},
  {Roegiers}  \& {Sbarufatti}}{{Kennea} et~al.}{2016}]{kennea2016gcn}
{Kennea} J.~A.,  {Lien} A.~Y.,  {Marshall} F.~E.,  {Palmer} D.~M.,  {Roegiers}
  T.~G.~R.,   {Sbarufatti} B.,  2016, GRB Coordinates Network, \href
  {https://ui.adsabs.harvard.edu/abs/2016GCN.19735....1K} {19735, 1}

\bibitem[\protect\citeauthoryear{{Kumar}, {Lu}  \& {Bhattacharya}}{{Kumar}
  et~al.}{2017}]{Kumar2017}
{Kumar} P.,  {Lu} W.,   {Bhattacharya} M.,  2017, \mn@doi [\mnras]
  {10.1093/mnras/stx665}, \href
  {https://ui.adsabs.harvard.edu/abs/2017MNRAS.468.2726K} {468, 2726}

\bibitem[\protect\citeauthoryear{{Li}, {Rea}, {Torres}  \& {de
  O{\~n}a-Wilhelmi}}{{Li} et~al.}{2017}]{Li2017}
{Li} J.,  {Rea} N.,  {Torres} D.~F.,   {de O{\~n}a-Wilhelmi} E.,  2017, \mn@doi
  [\apj] {10.3847/1538-4357/835/1/30}, \href
  {https://ui.adsabs.harvard.edu/abs/2017ApJ...835...30L} {835, 30}

\bibitem[\protect\citeauthoryear{{Li}, {Zrake}  \& {Beloborodov}}{{Li}
  et~al.}{2019}]{Li-Zrake2019}
{Li} X.,  {Zrake} J.,   {Beloborodov} A.~M.,  2019, \mn@doi [\apj]
  {10.3847/1538-4357/ab2a03}, \href
  {https://ui.adsabs.harvard.edu/abs/2019ApJ...881...13L} {881, 13}

\bibitem[\protect\citeauthoryear{{Li} et~al.,}{{Li} et~al.}{2020a}]{HXMT-FRB}
{Li} C.~K.,  et~al., 2020a, arXiv e-prints, \href
  {https://ui.adsabs.harvard.edu/abs/2020arXiv200511071L} {p. arXiv:2005.11071}

\bibitem[\protect\citeauthoryear{{Li}, {Yang}  \& {Dai}}{{Li}
  et~al.}{2020b}]{LiQiao-Chu2020}
{Li} Q.-C.,  {Yang} Y.-P.,   {Dai} Z.-G.,  2020b, \mn@doi [\apj]
  {10.3847/1538-4357/ab8db8}, \href
  {https://ui.adsabs.harvard.edu/abs/2020ApJ...896...71L} {896, 71}

\bibitem[\protect\citeauthoryear{{Lin} et~al.,}{{Lin} et~al.}{2020}]{FAST-FRB}
{Lin} L.,  et~al., 2020, arXiv e-prints, \href
  {https://ui.adsabs.harvard.edu/abs/2020arXiv200511479L} {p. arXiv:2005.11479}

\bibitem[\protect\citeauthoryear{{Lorimer}, {Bailes}, {McLaughlin}, {Narkevic}
  \& {Crawford}}{{Lorimer} et~al.}{2007}]{Lorimer2007}
{Lorimer} D.~R.,  {Bailes} M.,  {McLaughlin} M.~A.,  {Narkevic} D.~J.,
  {Crawford} F.,  2007, \mn@doi [Science] {10.1126/science.1147532}, \href
  {https://ui.adsabs.harvard.edu/abs/2007Sci...318..777L} {318, 777}

\bibitem[\protect\citeauthoryear{{Lu} \& {Kumar}}{{Lu} \&
  {Kumar}}{2016}]{Lu2016}
{Lu} W.,  {Kumar} P.,  2016, \mn@doi [\mnras] {10.1093/mnrasl/slw113}, \href
  {https://ui.adsabs.harvard.edu/abs/2016MNRAS.461L.122L} {461, L122}

\bibitem[\protect\citeauthoryear{{Lu}, {Kumar}  \& {Zhang}}{{Lu}
  et~al.}{2020}]{Lu2020}
{Lu} W.,  {Kumar} P.,   {Zhang} B.,  2020, arXiv e-prints, \href
  {https://ui.adsabs.harvard.edu/abs/2020arXiv200506736L} {p. arXiv:2005.06736}

\bibitem[\protect\citeauthoryear{{Luo et al.}}{{Luo et al.}}{2020}]{Luo2020}
{Luo et al.} R.,  2020, Nature, in press

\bibitem[\protect\citeauthoryear{{Lyubarsky}}{{Lyubarsky}}{2014}]{Lyubarsky2014}
{Lyubarsky} Y.,  2014, \mn@doi [\mnras] {10.1093/mnrasl/slu046}, \href
  {https://ui.adsabs.harvard.edu/abs/2014MNRAS.442L...9L} {442, L9}

\bibitem[\protect\citeauthoryear{{Margalit} \& {Metzger}}{{Margalit} \&
  {Metzger}}{2018}]{Margalit2018}
{Margalit} B.,  {Metzger} B.~D.,  2018, \mn@doi [\apjl]
  {10.3847/2041-8213/aaedad}, \href
  {https://ui.adsabs.harvard.edu/abs/2018ApJ...868L...4M} {868, L4}

\bibitem[\protect\citeauthoryear{{Mart{\'{\i}}n}, {Torres}  \&
  {Rea}}{{Mart{\'{\i}}n} et~al.}{2012}]{Martin2012}
{Mart{\'{\i}}n} J.,  {Torres} D.~F.,   {Rea} N.,  2012, \mn@doi [\mnras]
  {10.1111/j.1365-2966.2012.22014.x}, \href
  {http://adsabs.harvard.edu/abs/2012MNRAS.427..415M} {427, 415}

\bibitem[\protect\citeauthoryear{{Mart{\'{\i}}n}, {Torres}  \&
  {Pedaletti}}{{Mart{\'{\i}}n} et~al.}{2016}]{Martin2016}
{Mart{\'{\i}}n} J.,  {Torres} D.~F.,   {Pedaletti} G.,  2016, \mn@doi [\mnras]
  {10.1093/mnras/stw684}, \href
  {http://adsabs.harvard.edu/abs/2016MNRAS.459.3868M} {459, 3868}

\bibitem[\protect\citeauthoryear{{Mereghetti} et~al.,}{{Mereghetti}
  et~al.}{2020}]{Mereghetti2020}
{Mereghetti} S.,  et~al., 2020, \mn@doi [\apjl] {10.3847/2041-8213/aba2cf},
  \href {https://ui.adsabs.harvard.edu/abs/2020ApJ...898L..29M} {898, L29}

\bibitem[\protect\citeauthoryear{{Metzger}, {Berger}  \& {Margalit}}{{Metzger}
  et~al.}{2017}]{Metzger2017}
{Metzger} B.~D.,  {Berger} E.,   {Margalit} B.,  2017, \mn@doi [\apj]
  {10.3847/1538-4357/aa633d}, \href
  {http://adsabs.harvard.edu/abs/2017ApJ...841...14M} {841, 14}

\bibitem[\protect\citeauthoryear{{Metzger}, {Margalit}  \& {Sironi}}{{Metzger}
  et~al.}{2019}]{Metzger2019}
{Metzger} B.~D.,  {Margalit} B.,   {Sironi} L.,  2019, \mn@doi [\mnras]
  {10.1093/mnras/stz700}, \href
  {https://ui.adsabs.harvard.edu/abs/2019MNRAS.485.4091M} {485, 4091}

\bibitem[\protect\citeauthoryear{{Murase}, {Kashiyama}  \&
  {M{\'e}sz{\'a}ros}}{{Murase} et~al.}{2016}]{Murase2016}
{Murase} K.,  {Kashiyama} K.,   {M{\'e}sz{\'a}ros} P.,  2016, \mn@doi [\mnras]
  {10.1093/mnras/stw1328}, \href
  {https://ui.adsabs.harvard.edu/abs/2016MNRAS.461.1498M} {461, 1498}

\bibitem[\protect\citeauthoryear{{Nicholl}, {Williams}, {Berger}, {Villar},
  {Alexander}, {Eftekhari}  \& {Metzger}}{{Nicholl} et~al.}{2017}]{Nicholl2017}
{Nicholl} M.,  {Williams} P.~K.~G.,  {Berger} E.,  {Villar} V.~A.,  {Alexander}
  K.~D.,  {Eftekhari} T.,   {Metzger} B.~D.,  2017, \mn@doi [\apj]
  {10.3847/1538-4357/aa794d}, \href
  {https://ui.adsabs.harvard.edu/abs/2017ApJ...843...84N} {843, 84}

\bibitem[\protect\citeauthoryear{{Plotnikov} \& {Sironi}}{{Plotnikov} \&
  {Sironi}}{2019}]{Sironi2019}
{Plotnikov} I.,  {Sironi} L.,  2019, \mn@doi [\mnras] {10.1093/mnras/stz640},
  \href {https://ui.adsabs.harvard.edu/abs/2019MNRAS.485.3816P} {485, 3816}

\bibitem[\protect\citeauthoryear{{Popov} \& {Postnov}}{{Popov} \&
  {Postnov}}{2013}]{Popov2013}
{Popov} S.~B.,  {Postnov} K.~A.,  2013, arXiv e-prints, \href
  {https://ui.adsabs.harvard.edu/abs/2013arXiv1307.4924P} {p. arXiv:1307.4924}

\bibitem[\protect\citeauthoryear{{Rea} et~al.,}{{Rea} et~al.}{2010}]{Rea2010}
{Rea} N.,  et~al., 2010, \mn@doi [Science] {10.1126/science.1196088}, \href
  {https://ui.adsabs.harvard.edu/abs/2010Sci...330..944R} {330, 944}

\bibitem[\protect\citeauthoryear{{Rea}, {Pons}, {Torres}  \& {Turolla}}{{Rea}
  et~al.}{2012}]{Rea2012}
{Rea} N.,  {Pons} J.~A.,  {Torres} D.~F.,   {Turolla} R.,  2012, \mn@doi
  [\apjl] {10.1088/2041-8205/748/1/L12}, \href
  {http://adsabs.harvard.edu/abs/2012ApJ...748L..12R} {748, L12}

\bibitem[\protect\citeauthoryear{{Rea}, {Vigan{\`o}}, {Israel}, {Pons}  \&
  {Torres}}{{Rea} et~al.}{2014}]{Rea2014}
{Rea} N.,  {Vigan{\`o}} D.,  {Israel} G.~L.,  {Pons} J.~A.,   {Torres} D.~F.,
  2014, \mn@doi [\apjl] {10.1088/2041-8205/781/1/L17}, \href
  {http://adsabs.harvard.edu/abs/2014ApJ...781L..17R} {781, L17}

\bibitem[\protect\citeauthoryear{{Safi-Harb} \& {Kumar}}{{Safi-Harb} \&
  {Kumar}}{2008}]{safiharb2008}
{Safi-Harb} S.,  {Kumar} H.~S.,  2008, \mn@doi [\apj] {10.1086/590359}, \href
  {https://ui.adsabs.harvard.edu/abs/2008ApJ...684..532S} {684, 532}

\bibitem[\protect\citeauthoryear{{The CHIME/FRB Collaboration} et~al.,}{{The
  CHIME/FRB Collaboration} et~al.}{2020}]{CHIME}
{The CHIME/FRB Collaboration} et~al., 2020, arXiv e-prints, \href
  {https://ui.adsabs.harvard.edu/abs/2020arXiv200510324T} {p. arXiv:2005.10324}

\bibitem[\protect\citeauthoryear{{Thompson} \& {Blaes}}{{Thompson} \&
  {Blaes}}{1998}]{Thompson1998}
{Thompson} C.,  {Blaes} O.,  1998, \mn@doi [\prd] {10.1103/PhysRevD.57.3219},
  \href {https://ui.adsabs.harvard.edu/abs/1998PhRvD..57.3219T} {57, 3219}

\bibitem[\protect\citeauthoryear{{Thompson} \& {Duncan}}{{Thompson} \&
  {Duncan}}{1996}]{Thompson1996}
{Thompson} C.,  {Duncan} R.~C.,  1996, \mn@doi [\apj] {10.1086/178147}, \href
  {https://ui.adsabs.harvard.edu/abs/1996ApJ...473..322T} {473, 322}

\bibitem[\protect\citeauthoryear{{Torres}}{{Torres}}{2017}]{Torres2017}
{Torres} D.~F.,  2017, \mn@doi [\apj] {10.3847/1538-4357/835/1/54}, \href
  {http://adsabs.harvard.edu/abs/2017ApJ...835...54T} {835, 54}

\bibitem[\protect\citeauthoryear{{Torres}, {Cillis}, {Mart{\'{\i}}n}  \& {de
  O{\~n}a Wilhelmi}}{{Torres} et~al.}{2014}]{Torres2014}
{Torres} D.~F.,  {Cillis} A.,  {Mart{\'{\i}}n} J.,   {de O{\~n}a Wilhelmi} E.,
  2014, \mn@doi [Journal of High Energy Astrophysics]
  {10.1016/j.jheap.2014.02.001}, \href
  {http://adsabs.harvard.edu/abs/2014JHEAp...1...31T} {1, 31}

\bibitem[\protect\citeauthoryear{{Turolla}, {Zane}  \& {Watts}}{{Turolla}
  et~al.}{2015}]{turolla2015}
{Turolla} R.,  {Zane} S.,   {Watts} A.~L.,  2015, \mn@doi [Reports on Progress
  in Physics] {10.1088/0034-4885/78/11/116901}, \href
  {https://ui.adsabs.harvard.edu/abs/2015RPPh...78k6901T} {78, 116901}

\bibitem[\protect\citeauthoryear{{Yang} \& {Dai}}{{Yang} \&
  {Dai}}{2019}]{Yang2019}
{Yang} Y.-H.,  {Dai} Z.-G.,  2019, \mn@doi [\apj] {10.3847/1538-4357/ab48dd},
  \href {https://ui.adsabs.harvard.edu/abs/2019ApJ...885..149Y} {885, 149}

\bibitem[\protect\citeauthoryear{{Yang} \& {Zhang}}{{Yang} \&
  {Zhang}}{2018}]{YangZhang2018}
{Yang} Y.-P.,  {Zhang} B.,  2018, \mn@doi [\apj] {10.3847/1538-4357/aae685},
  \href {https://ui.adsabs.harvard.edu/abs/2018ApJ...868...31Y} {868, 31}

\bibitem[\protect\citeauthoryear{{Yang}, {Zhang}  \& {Dai}}{{Yang}
  et~al.}{2016}]{Yang2016}
{Yang} Y.-P.,  {Zhang} B.,   {Dai} Z.-G.,  2016, \mn@doi [\apjl]
  {10.3847/2041-8205/819/1/L12}, \href
  {https://ui.adsabs.harvard.edu/abs/2016ApJ...819L..12Y} {819, L12}

\bibitem[\protect\citeauthoryear{{Younes} et~al.,}{{Younes}
  et~al.}{2016a}]{Younes2016}
{Younes} G.,  et~al., 2016a, \mn@doi [\apj] {10.3847/0004-637X/824/2/138},
  \href {http://adsabs.harvard.edu/abs/2016ApJ...824..138Y} {824, 138}

\bibitem[\protect\citeauthoryear{{Younes}, {Archibald}, {Kouveliotou}, {Kaspi},
  {Ray}, {McEnery}  \& {Fermi LAT Collaboration}}{{Younes}
  et~al.}{2016b}]{Younes2016gcn}
{Younes} G.,  {Archibald} R.,  {Kouveliotou} C.,  {Kaspi} V.,  {Ray} P.~S.,
  {McEnery} J.,   {Fermi LAT Collaboration} 2016b, The Astronomer's Telegram,
  \href {https://ui.adsabs.harvard.edu/abs/2016ATel.9378....1Y} {9378, 1}

\bibitem[\protect\citeauthoryear{{Zhang}}{{Zhang}}{2020}]{Zhang2020}
{Zhang} R.,  2020, Nature, in press

\bibitem[\protect\citeauthoryear{{de Jager} \& {Djannati-Ata{\"\i}}}{{de Jager}
  \& {Djannati-Ata{\"\i}}}{2009}]{deJager2009}
{de Jager} O.~C.,  {Djannati-Ata{\"\i}} A.,  2009, in {Becker} W.,  ed.,
  Astrophysics and Space Science Library Vol. 357, Astrophysics and Space
  Science Library. p.~451 (\mn@eprint {arXiv} {0803.0116}),
  \mn@doi{10.1007/978-3-540-76965-1_17}

\bibitem[\protect\citeauthoryear{{de Jager}, {Harding}, {Michelson}, {Nel},
  {Nolan}, {Sreekumar}  \& {Thompson}}{{de Jager} et~al.}{1996}]{deJager1996}
{de Jager} O.~C.,  {Harding} A.~K.,  {Michelson} P.~F.,  {Nel} H.~I.,  {Nolan}
  P.~L.,  {Sreekumar} P.,   {Thompson} D.~J.,  1996, \mn@doi [\apj]
  {10.1086/176726}, \href {http://adsabs.harvard.edu/abs/1996ApJ...457..253D}
  {457, 253}

\makeatother
\end{thebibliography}

\label{lastpage}
\end{document}